# Two Orders of Magnitude Enhancement in Oxide Ion Conductivity in $Cu_2P_2O_7$ via Vanadium Substitution: A Pathway Toward SOFC Electrolytes

Bibhas Ghanta[1,2], Kuldeep Singh Chikara[1,2], Uttam Kumar Goutam[3], Anup Kumar Bera[1,2,*], and Seikh Mohammad Yusuf[1,2]

*[1]Solid State Physics Division, Bhabha Atomic Research Centre, Mumbai, 400085, India*
*[2]Homi Bhabha National Institute, Anushaktinagar, Mumbai 400094, India*
*[3]Technical Physics Division, Bhabha Atomic Research Centre, Mumbai, 400085, India*
*[*] E-mail: akbera@barc.gov.in*

**ABSTRACT:**

In the quest of green energy, Solid Oxide Fuel Cells (SOFC) have drawn considerable attention for chemical-to-electric energy conversion. Electrolyte material with high oxygen ionic conductivity is essential for an efficient SOFC operation. However, materials with higher oxygen ionic conduction are rear in nature. Selective chemical substitution is an effective route to enhance ionic conductivity of an existing material. In the present paper, we report an enhancement of ionic conductivity in $Cu_2P_{2-x}V_xO_7$ by vanadium substitution. The electrical (dc and ac conductivity, diffusivity, hopping rate, electric modulus and dielectric properties) and crystal structural properties of $Cu_2P_{2-x}V_xO_7$ ($x$ = 0, 0.4, 0.6, 0.8 and 1) are investigated by impedance spectroscopy and neutron diffraction, respectively. X-ray photoelectron spectroscopy (XPS) study confirms the presence of $Cu^{2+}$, $P^{5+}$and $V^{5+}$ mono-valence states. The dc conductivity results reveal a two orders of magnitude enhancement of ionic conductivity from ~$3.81\times10^{-5}$ S cm$^{-1}$ for $x$ =0 to ~$2.08\times10^{-3}$ S cm$^{-1}$ for $x$ =1 at 993 K, revealing a possible application in SOFCs. DC transport number studies reveal that the total conductivity is dominated by ionic conduction (> 95%). In addition, the diffusivity and hopping rate of oxide ions increase with increasing $x$. Besides, ac conductivity, electric modulus and dielectric properties have been investigated to illustrate the microscopic conduction mechanism. The derived results suggest that the mechanism for ionic conduction is the correlated barrier hopping (CBH) process. The soft-bond valence sum (BVS) analysis of the neutron diffraction patterns reveals the three-dimensional (3D) oxide ion conduction pathways within the crystal structure. The crystal structural study reveals an increase in unit cell volume, polyhedra volume and polyhedral distortion with increasing $x$, which lead to an enhancement of the ionic conduction. The present study provides a pathway to enhance the ionic conductivity, as well as understanding of microscopic conduction mechanism, ionic conduction pathways and the role of crystal structure on the ionic conduction.

**KEYWORDS:** *pyrophosphate, impedance spectroscopy, neutron diffraction, SOFC, correlated barrier hopping (CBH), BVS analysis*

## 1. INTRODUCTION

Green energy sources are naturally replenished, making them a sustainable choice for long-term energy needs. In the thrust of green energy, the Solid Oxide Fuel Cells (SOFC)[1] are attracted a considerable attraction in recent years due to their several salient features including high efficiency, low emissions (minimal production of pollutants), silent-operation (no noise), long-lasting and simplicity. Efficiency of the SOFC, which convert chemical energy to electrical energy, highly depends on the oxide-ion conductivity of electrolyte[1, 2]. Presently, high operating temperature (≥ 1073 K) is the primary limitation for SOFC[3]. Electrolytes with oxygen ionic conductivity higher than $10^{-2}$ S cm$^{-1}$ at low temperature (≤ 500 K) are favorable for a cost-effective SOFC[3, 4]. Although, significant studies on ionic conductivity in oxide electrolytes have been reported in the literature[4-6], suitable material for oxide electrolytes still remains elusive. Among the oxide electrolytes, yttria-stabilized zirconia (YSZ-$Zr_{0.92}Y_{0.16}O_{2.08}$, $\sigma_{dc}$ ~ $10^{-2}$ S cm$^{-1}$ at ~ 1073 K) is mostly being used in commercial SOFC[7]. However, due to its high operating temperature, the associated cost for its operation remains significantly high, and efficiency reduces rapidly. To reduce the operating costs and degradation rates, considerable studies are being conducted to lower the working temperature either by material engineering or by discovering new materials [3, 7, 8]. In present days, research in solid oxide electrolytes primarily focuses to achieve the desired value of ionic conductivity (≥ $10^{-2}$ S cm$^{-1}$) at considerably lower temperature, preferably below ~500 K.

Recently, pyrophosphate compounds have attracted a lot of attention for solid electrolytes and ionic conductors[9-13] . Among these pyrophosphates, the divalent-pyrophosphate compounds fall in a family with the general formula $M_2P_2O_7$, where $M$ = divalent alkaline or transition metal cations ($M$ = Mg, Ca, Sr, Ba, Mn, Cu, Ni, Zn, Co, Cd, Pb). These compounds appear to be crystalized in two distinct crystal structure conformations (thortveitite and dichromate) which are characterized by the values of ionic radius (IR) of $M$ cation[14, 15]. For the thortveitite structure, radius of $M$ cation ($M$ = Mg, Mn, Cu, Ni, Zn, Co) is less than 0.97 Å. On the other hand, for the dichromate structure, radius of $M$ cation ($M$ = Ca, Sr, Ba, Pb, Cd) is greater than 0.97 Å. The $M_2P_2O_7$ compounds are low-cost, nontoxicity and eco-friendly materials, having good conductivity, chemical and thermal stability[16] as well as easy to synthesis. The $M_2P_2O_7$ compounds show a crystal structural phase transition from low-temperature $\alpha$-phase to high-temperature $\beta$-phase. In this $M_2P_2O_7$ family, the transition metal-based compound copper pyrophosphate $Cu_2P_2O_7$, is of our present interest. As a prospective material, the compound $Cu_2P_2O_7$ has attracted much attention due to its some intriguing and anomalous properties as well as applications, including strong negative thermal expansion, electrochemical, sensors and energy storage devices[17-27]. The compound $Cu_2P_2O_7$ has excellent reversible crystal structural property[16]. The $Cu_2P_2O_7$ is nonhygroscopic and stable in air up to 1443 K, where after it decomposes into $Cu_3(PO_4)_2$ and $Cu^0$. While in argon atmosphere, the material does not decompose up to 1473 K[28]. The compound with good crystal structural stability offers a long-term durability, and therefore, can maintain stable ionic conductivity over a long-period of operation of a real SOFC cell.

The compound $Cu_2P_2O_7$ with thortveitite crystal structure (Monoclinic, space group: $C2/c$) exhibits a stable phase ($\alpha$-phase) at room temperature and undergoes a reversible phase transition to $\beta$-phase (Monoclinic, space group: $C2/m$) at 350 K[17, 29]. The key difference between $\alpha$-phase and $\beta$-phase is the positional disorder of the bridging O1 atom in the P–O1–P bridge of $[P_2O_7]$ unit in the $\beta$-phase. Although $Cu_2P_2O_7$ reveals a strong crystal structural stability, the oxygen ionic conductivity of $Cu_2P_2O_7$ is reported to be low ~$1.14\times10^{-8}$ S $cm^{-1}$ at 300 K[11], which is not suitable for SOFC. Therefore, it is necessary to find ways to enhance the ionic conductivity of $Cu_2P_2O_7$ for its possible applications in SOFC. Further, the nature of temperature dependence of the ionic conductivity as well as underlying mechanism of ionic conduction is unknown, which demands a detailed temperature dependent ionic conductivity study. An effective route to enhance the ionic conductivity of $Cu_2P_2O_7$ is V-substitution at P site. The reported results reveal that the solubility of V in $Cu_2P_{2-x}V_xO_7$ is 100% where the substituted compounds retain a monoclinic phase (*C2/c*) up to $x$ ~ 1.5 and then transformed to orthorhombic phase (*Fdd2*) without appearance of any secondary phase[27]. The temperature-dependent high-resolution synchrotron x-ray diffraction study on the $x$ =1 ($Cu_2PVO_7$) compound reported that the compound is thermally stable with monoclinic phase up to the highest measured temperature of 773 K[27].

The present study aims to synthesize the compounds $Cu_2P_{2-x}V_xO_7$ ($x$ = 0, 0.4, 0.6, 0.8 and 1) using the solid-state reaction method and investigate their ionic conduction, electrical and crystal structural properties by impedance spectroscopy and x-ray/neutron diffraction, respectively. To maintain structural consistency with the parent compound $Cu_2P_2O_7$ ($x$ = 0), and to enable a reliable and systematic comparison of conductivity data within the same monoclinic structural framework, we, have restricted our study over the V substitution range $x$ = 0-1. The XPS and DC transport number measurements were employed to determine the valence states of the cations, and the contributions of ionic and electronic conductivity to the total conductivity, respectively. Our detailed analysis shows a two orders of magnitude enhancement of the ionic conductivity in $Cu_2P_2O_7$ by a larger $V^{5+}$-substitution at the $P^{5+}$ site. At the same time, a decrease of activation energy ($E_a$) from 1.47(2) eV for $x$ =0 to 1.31(2) eV for $x$ =1 is achieved. Further, we have determined the diffusivity, hopping rate of oxide ions, as well as microscopic mechanism of ionic conduction. The results reveal the correlated barrier hopping (CBH) process for ionic conduction mechanism. The analysis of XPS spectra confirms mono-valance state for Cu, P and V ions, i.e., $Cu^{2+}$, $P^{5+}$ and $V^{5+}$. The DC transport number measurements reveal that the electronic transference number is negligible (below 5%), confirming that the observed conductivity predominantly arises from the ionic contribution (> 95%). The oxide ion conduction pathways within the crystal structure are illustrated by a comprehensive soft-bond valence sum (BVS) analysis of neutron diffraction patterns. Present work, thus, provides a potential route to increase the oxide ionic conductivity of $Cu_2P_2O_7$, as well as understanding of the microscopic conduction mechanism and the determination of ionic conduction pathways, which are useful for the advancement of SOFC devices.

## 2. EXPERIMENTAL DETAILS

The powder samples of copper pyrophosphate compounds ($Cu_2P_{2-x}V_xO_7$, $x$ = 0, 0.4, 0.6, 0.8 and 1) were synthesized by the solid-state reaction method. The stoichiometric mixtures (as per required compositions) of high purity (Sigma-Aldrich, > 99.99%) precursors of CuO, $V_2O_5$ and $(NH_4)_2HPO_4$ were well ground using agate mortar and pestle. In the first stage, mixed powders were calcinated at 573 K for 6 hours to remove $NH_3$ and $H_2O$. In the second stage, the calcined powders were ground again and pelletized, and then sintered at 1073 K for 24 hours.

The phase purity of synthesized polycrystalline compounds of $Cu_2P_{2-x}V_xO_7$ ($x$ = 0, 0.4, 0.6, 0.8 and 1) has been confirmed by both x-ray diffraction (XRD) and neutron powder diffraction (NPD) at room temperature. The x-ray diffraction measurements were carried out over a scattering angular range of $10° \leq 2\theta \leq 70°$ with a scan step size of 0.02° using a laboratory-based x-ray diffractometer (make: Rigaku, Japan) with a standard Cu $K_\alpha$ source. The neutron diffraction measurements were carried out using the PD-1 diffractometer[30] ($\lambda$= 1.094 Å), comprised of three linear position sensitive neutron detectors covering an angular range of 4°−70°, at Dhruva Research Reactor, BARC, Mumbai. The powders samples were filled inside an 8 mm cylindrical vanadium-can. The neutron diffraction pattern was recorded for around 12 hours. The analysis of powder diffraction patterns was carried out by the Rietveld refinement method with the help of the Fullprof computer program[31]. The soft-bond valence sum (BVS) analysis was performed using the Bond_STR program[32], an inbuilt package in Fullprof computer program. The migration pathways were plotted using the VESTA software[33]. The x-ray photoelectron spectroscopy (XPS) measurements were carried out for all the compounds $Cu_2P_{2-x}V_xO_7$ ($x$ = 0, 0.4, 0.6, 0.8 and 1) using the synchrotron radiation at BL-14 beam line, Indus 2 at Synchrotron Radiation Facility, RRCAT, India.

For the impedance measurements, fine powders of $Cu_2P_{2-x}V_xO_7$ were pressed into disc shaped pellets (10 mm diameter) under 10 Ton pressure using a hydraulic press machine. Further, the pellets were isothermally sintered at 1073 K for 6 hours to increase the density above 90% of the theoretical density. The flat surfaces of pellets were uniformly coated with gold paste and dried at 573 K for 3 hours before performing measurements. Impedance measurements were performed using a PSM1735−NumetriQ impedance analyzer (Newtons4th Ltd., U.K.) with an ac electric field amplitude of 0.4 V and in the frequency ($f$) range of 1–$10^7$ Hz, over a temperature range of 423–993 K in air. The EIS Spectrum Analyser software[34] was used for the impedance data analysis. The measured impedance data were corrected by employing the density correction following a procedure as reported in literature[35]. The DC transport number measurements[36] were carried out at room temperature on the representative $x$ = 0, 0.8 and 1 compounds, under a fixed applied dc voltage (0.5 V for $x$ = 0; 0.2 V for $x$ = 0.8 and 1, respectively) across the pellets using the ion-blocking gold electrodes. The data were recorded using the BioLogic make SP-300 Potentiostat.

## 3. RESULTS

### 3.1. Crystal Structural Properties

The crystal structural properties of $Cu_2P_{2-x}V_xO_7$ ($x$ = 0, 0.4, 0.6, 0.8 and 1) compounds have been investigated by combined XRD and NPD studies. The XRD and NPD patterns for all the compounds are shown in Figures 1a and 1c−g, respectively. The diffraction patterns for all the compounds $x$ = 0−1 have been indexed corresponding to the standard data of ICSD_CollCode14369 and ICSD_CollCode121992, reporting the same monoclinic crystal structure with space group $C2/c$. The Rietveld analysis of the XRD and NPD patterns reveals that all the compounds belong to the $\alpha$-phase with Monoclinic symmetry (space group: $C2/c$)[27, 29]. The refined values of crystal structural parameters and unit cell volumes are listed in Table 1. The analyses confirm that the vanadium (V) ions are substituted at the P site. The confirmation of V-substitution at the P site is evident from the shifting of Bragg peaks towards lower value of momentum transfer (Q) with increasing the substitution. The shifting of Bragg peaks reveals an increase of lattice constants and unit cell volume with the substitution of larger $V^{5+}$ ions (IR: 0.35 Å) at the position of smaller $P^{5+}$ ions (IR: 0.17 Å). A large shifting of nuclear Bragg peak (-202) towards a lower Q value is observed. The detailed composition-dependent shiftings of some of the observed Bragg peaks are shown in Figures 1b and 1h, respectively, for XRD and NPD patterns. Detailed analyses reveal that with the V-substitution, the lattice constants *a* and *c* increase linearly, whereas lattice constant *b* slightly decreases [Figure 2a]. It is also observed that the unit cell volume (*V*) and monoclinic angle $\beta$ increase linearly with increasing the concentration (*x*) of vanadium [Figure 2b]. The normalized changes of *a*, *b*, *c*, $\beta$ and *V* with respect to $x = 0$ compound are shown in Figure 2c. The observed changes in the unit cell volume *V* for $x = 0.4$, 0.6, 0.8 and 1 are ~3.90 %, 6.04 %, 8.06 % and 10.43 % with reference to the $x = 0$ compound.

A schematic of crystal structure is shown in Figure 3a. Within the unit cell, there are eight Cu atoms, eight *B* (P or P/V) atoms, and twenty-eight O atoms. Cu atoms occupy the 8*f* site, *B* atoms occupy the 8*f* site, and O atoms occupy the 8*f* and 4*e* sites. It is mentioned that Cu, *B*, and O atoms each occupy symmetrically-distinct 8*f* sites, listed in Table 1. The unit cell consists of two different polyhedral units, viz., $CuO_5$ pyramidal and $BO_4$ ($B$ = P, V) tetrahedral units. The Cu atom coordinates with five O atoms (two O2, two O3, and one O4) and forms $CuO_5$ pyramid, and the *B*-site atom coordinates with four O atoms (O1, O2, O3, and O4) to form $BO_4$ tetrahedron [Figure 3b]. The $CuO_5$ pyramids and $BO_4$ tetrahedra make alternating layers along the crystallographic *c*-axis. Within a given layer (*ab* plane), the edge-shared $CuO_5$ units (sharing their O2-O2 and O3-O3 edges) make a zigzag chain along the diagonal direction in the same plane. The arrangements of the zigzag chains are along opposite diagonal directions in the adjacent layers. Adjacent $CuO_5$ layers are linked by corner-shared $B_2O_7$ groups (sharing O2, O3, and O4 atoms at the corners) which are formed by two $BO_4$ tetrahedral units sharing O1 atoms at the vertex (making *B*-O1-*B* bridge) [Figure 3b]. The substitution of larger $V^{5+}$ ion at the $P^{5+}$ site reduces the rigidity of the polyhedra by increasing $BO_4$ polyhedra volume and $BO_4$ average bond length as well as unit cell volume. The values of unit cell volume, $BO_4$ polyhedra volume, and $BO_4$ average bond length increase by 10.43 %, 20 % and 6 %, respectively, for vanadium substitution of $x = 1$. Additionally, angular distortion in $CuO_5$ and stretching distortion in $BO_4$ increase with the V-substitution. The increase in unit cell and polyhedral volumes is expected to facilitate easier ion migration. An increase in bond length reduces the bond strength, which is expected to make easier for oxide ions conduction. An enhancement of oxygen ionic conductivity is, therefore, expected with the V-substitution in the present study as discussed later.

### 3.2. Valence State of Cations

To further examine the valence states of cations in $Cu_2P_{2-x}V_xO_7$ ($x$ = 0, 0.4, 0.6, 0.8 and 1), *i.e.*, valence states of V, Cu, and P, XPS spectra were recorded for all the compounds. The V $2p_{3/2}$ spectra are shown in Figures 3d−g, revealing V $2p_{3/2}$ peak with binding energy of ~517 eV, confirming the $V^{5+}$ valence sate[37]. The spectra of Cu 3*s* and P 2*p* are shown in Figures 3h−l. The spectra are fitted with Cu 3*s* (main + satellite) and P 2*p* peaks. The Cu 3*s* and P 2*p* peaks with binding energy of ~125 eV and ~ 133 eV correspond to the $Cu^{2+}$ and $P^{5+}$ valence sates[38,39], respectively. The Cu $2p_{3/2}$ spectra are shown in Figures 3m−q. The Cu $2p_{3/2}$ spectra were fitted with $Cu^{2+}$ multiplet peaks. The analyses reveal a main peak with binding energy of ~935.2 eV accompanied by the characteristic shakeup satellite peaks (938−945 eV). The binding energy of ~935.2 eV corresponding to the $Cu^{2+}$ valence sate[40, 38]. Therefore, the XPS results confirm the presence of $Cu^{2+}$, $P^{5+}$ and $V^{5+}$ valence states, i.e., mono-valence cations in all the $Cu_2P_{2-x}V_xO_7$ ($x$ = 0, 0.4, 0.6, 0.8 and 1) compounds.

### 3.3. Electrical Conduction Properties

#### 3.3.1 Dc Conductivity

The real part $Z'$ and imaginary part $Z''$ of the complex impedance $Z^* = Z' - jZ''$ for all the compounds ($x$ = 0, 0.4, 0.6, 0.8, and 1), measured at various temperatures ($T$) as a function of frequency ($f$), are shown in Figures 4a−e and 4f−j, respectively. For $x$ = 0.6 and 0.8, a non-monotonic behavior has been found in the temperature dependent $Z'$ value. With increasing temperature, $Z'$ value first decreases gradually up to 603 K, then increases with further increasing temperature up to ~ 653 K (for $x$ = 0.6) and ~ 633 K (for $x$ = 0.8), and then decreases again up to highest measured temperature of 993 K. On the other hand, for the compounds with $x$ = 0, 0.4, and 1, with increasing temperature, $Z'$ value decreases monotonically over the whole temperature range up to 993 K with a change in the rate around the intermediate temperature range. The overall decreasing trend of $Z'$ with increasing temperature (except for the $x$ = 0.6 and 0.8 compounds over the intermediate temperature range) suggests a negative temperature coefficient of resistance, hence, a semi-conducting behavior. The non-monotonic behavior of resistance in the intermediate temperature range is observed around the crystal structural phase transition (from the $\alpha$ to the $\beta$-phase) temperature, where resistance increases with increasing the temperature. Such a change in the temperature coefficient of resistance from -ve to +ve with increasing temperature was also reported for $BaTiO_3$ and Ti doped $SrFeO_3$ ceramic compounds[41, 42] around the crystal structural phase transition temperature and the positive coefficient of the resistance is reported due to charge trapping at the acceptor states around the phase transition temperature. As the trapped charges are not available for conduction, an increase in the resistivity is found. A similar charge trapping phenomena may be responsible for the increase of the resistance in the intermediate temperature range for the studied compounds. For all the present compounds, each of the $Z''$ ($f$) curves exhibits a peak at a characteristic frequency $f_Z$ at all measured temperatures. The existence of peak indicates a presence of relaxation phenomena. The peak in the $Z''$ curve shifts towards higher frequencies with increasing temperature. The shifting of the peak frequency towards the higher frequencies with increasing temperature reveals a decrease in the relaxation time ($\tau_Z$), hence a faster ionic conduction. The presence of a distribution of relaxation time in the studied materials (as evident from the board peaks) reveals the non-Debye type relaxation process. In real materials like the studied compounds, the distribution of relaxation time has been found due to the presence of non-uniform grain-size distribution, non-uniform grain and grain boundary-orientation, and crystal structural disorder. Relaxation time ($\tau_Z$) can be calculated using the formula:

$$\tau_Z = \frac{1}{2\pi f_Z} \qquad (1)$$

where, $f_Z$ is the characteristic frequency at which the $Z''$ reaches its maximum value. The value of $\tau_Z$ decreases with increasing the temperature to 993 K. An opposite behavior has been observed over the 603−653 K (for $x$ = 0.6) and 603−633 K (for $x$ = 0.8) around the crystal structural phase transition temperature corresponding to the $\alpha$ to $\beta$-phase transition.

The Cole-Cole (-$Z''$ vs $Z'$) plots[43, 44] are shown in Figures 5a−g, for all the compounds at selected temperatures. The experimental data were best fitted with an equivalent circuit (based on the brick-layer model[45]) [Figure 5i] over the whole measured temperature range up to 993 K. The equivalent circuit consists of two sub-circuits connected in series. The first subcircuit contains resistance ($R_g$) and a constant phase element ($CPE_g$) in parallel, representing the grain contribution to the ionic conductivity[32]. The second subcircuit contains a constant phase element ($CPE_{gb}$) alongside the parallel $R_{gb}$-$C_{gb}$ elements to account for the grain boundary contribution. The presence of a constant phase element (*CPE*)[44] reveals the deviation from the ideal Debye-type behavior and the existence of a distribution of the relaxation time. The impedance of *CPE* in the complex plane is defined as:

$$Z_{CPE} = \frac{1}{Q\,(j\omega)^p} \tag{2}$$

where, $Q$ is the value of capacitance of *CPE*, and $p$ is the fractal exponent ranges between 0 and 1. When $p = 0$ and 1, *CPE* behaves as an ideal resistor and ideal capacitor[46], respectively.

The total dc conductivity $\sigma_{dc}$ was calculated from the following relation:

$$\sigma_{dc} = \frac{t}{A\,(R_g + R_{gb})} = \frac{t}{A\,R_{dc}} \tag{3}$$

where, $A$, $t$ and $R_{dc}$ represent the area of the cross-section, thickness and total resistance of the pellet, respectively. Total resistance $R_{dc}$, which is equivalent to the sum of the individual's contribution of grain and grain boundary resistance *i.e.*, $R_{dc} = R_g + R_{gb}$,, was obtained from the fitting of the Cole-Cole plots and corresponding to the intersect of the extrapolated Cole-Cole curve to the $Z'$-axis (X-axis) at the zero frequency. Figure 6a shows the temperature dependence of the total dc conductivity ($\sigma_{dc}$) derived using Eq. (3) for all the compounds. The Arrhenius plots [$\sigma_{dc}(T) = (\sigma_0/T)\, exp\,(-E_a/k_BT)$] of the $\sigma_{dc}$ curves reveal straight lines over lower and higher temperature regions with anomalies around the structural phase transition temperature of the respective compounds [Figures 6b−f]. In the Arrhenius relation, $\sigma_0$ is the pre-exponential factor, and it depends upon the concentration of mobile charge carriers[47, 48], $E_a$ is the activation energy for ionic conduction, and $k_B$ is the Boltzmann constant. Activation energy $E_a$ and pre-exponential factor $\sigma_0$ are determined from the linear fits [Figures 6b−f]. The activation energies $E_a$ lie in the range of 0.43−1.04 eV for the α-phase and in the range of 1.19−1.47 eV for the $\beta$-phase. The value of dc conductivity increases with increasing temperature. Such temperature dependence behaviour of dc conductivity confirms the thermally activated ionic conduction. It is found that all the $\sigma_{dc}$, $E_a$ and $\sigma_0$ parameters show a strong dependence on the substitution of V-content. The variations of $E_a$ and $\sigma_0$ with composition are shown in Figures 6g and 6h, respectively. In the α-phase, both $E_a$ and $\sigma_0$ steadily increase with increasing the $x$-value. On the other hand, in the $\beta$-phase, both $E_a$ and $\sigma_0$ values remain almost constant. It is noticeable that for a particular composition, the value of $\sigma_0$ is higher in the $\beta$-phase as compared to that of the $\alpha$-phase which indicates that the increase in mobile charge carrier concentration leading to an increase in dc conductivity. Most important result is that in the $\beta$-phase, a two orders of magnitude enhancement of conductivity has been found with the 50% V-substitution, i.e., $x = 1$. We discuss below the mechanism of ionic conduction and the possible origin of the enhancement of ionic conductivity with the V-substitution.

### 3.3.2 DC Transport Number

The relative contributions of ionic and electronic conductions have been measured using the DC transport number measurements on the representative compounds $x = 0$, 0.8 and 1. In this technique, a fixed DC voltage is applied across the sample and the DC current is monitored as a function of time. The results of these measurements are shown in Figure 5h. It is noticeable that, initially, the current decreases rapidly with time due to the blocking of the ionic species by the blocking electrode and then attains a constant value (corresponding to the electronic current $i_e$) under fully blocked condition. The electronic transference numbers ($t_e$) = $i_e/i_t$, where $i_t$ is initial/total current, are determined to be 0.0295 for $x$ = 0, 0.0453 for $x$ = 0.8 and 0.0456 for $x$ = 1, respectively. The corresponding ionic transference numbers ($t_i = 1 - t_e$) are 0.9705 for $x$ = 0, 0.9547 for $x$ = 0.8 and 0.9544 for $x$ = 1, respectively. Therefore, the above results corroborate that conductivity in the studied compounds $Cu_2P_{2-x}V_xO_7$ arises predominantly (> 95%) from ionic conduction. The low fraction of electronic conduction (<5%) is also supported by the DFT results of $Cu_2P_2O_7$ which predict an insulating state with a large band gap of $E_g$ = 3.17 eV[26]. Experimentally the electronic band gap value is determined to be $E_g$ = 3.95 eV[49] which is not favorable for electronic conduction.

### 3.3.3 Diffusivity and Hopping Rate

The diffusion constant (diffusivity) $D$ of oxide ions conduction has been determined from the dc conductivity data for the two end compounds with $x$ = 0 and 1. The diffusion constant $D$ is derived based on the Nernst−Einstein relationship, as used for other compound [Ref. 33].:

$$D = \frac{\sigma_{dc}\, k_B T}{N q^2} \qquad (4)$$

where, $N$ is the concentration of oxide ions, $q$ is the ionic charge of oxide ion ($q$ = -2e for oxygen ion) and $\sigma_{dc}$ is the dc conductivity. The temperature dependence of $D$ values for the $x$ = 0 and 1 compounds are shown in Figure 6i. The value of $D$ increases with increasing temperature for both the compounds with $x$ = 0 and 1 in the $\beta$-phase. Whereas, the value of $D$ for the $x$ = 1 compound (~$6.49\times10^{-11}$ cm$^2$/sec at 773 K) is two orders of magnitude higher than that of the $x$ = 0 compound (~$6.11\times10^{-13}$ cm$^2$/sec at 773 K) over the full temperature range in the $\beta$-phase [Figure 6i]. Furthermore, the hopping rate ($\Delta\tau$), defined by the time taken by an ion to hop from initial site to the final site, of mobile oxide ions was determined using the equation[33]:

$$\Delta\tau = \frac{< l^2 >}{2\, d\, D} \qquad (5)$$

where, $< l^2 >$ is the mean square jump length (hopping distance) of mobile oxide ions. The $< l^2 >$ values (an average distance between the neighboring oxygen ion sites) have been calculated from the crystal structural parameters obtained from Rietveld analyses of x-ray and neutron diffraction patterns. The $d$ is the dimensionality of oxide ion diffusion, For the studied compounds, $d$ =3, due to the three-dimensional oxide ion migration (a detailed discussion is given in Section 3.2.6). The temperature dependent hopping rate curves for the $x$ = 0 and 1 compounds are shown in Figure 6j. For both the compounds (*i.e.*, $x$ = 0 and 1), the hopping rate decreases with increasing temperature. Moreover, it is observed that value of $\Delta\tau$ for the $x$ = 1 is two orders of magnitude smaller than that of the $x$ = 0 compound, suggesting a faster ionic conduction for the compound with $x$ = 1 [Figure 6j].

### 3.3.4 Ac Conductivity

The ionic conduction mechanism for the studied compounds $Cu_2P_{2-x}V_xO_7$ ($x$ = 0, 0.4, 0.6, 0.8 and 1) has been investigated by frequency ($f$) dependent ac conductivity. The complex conductivity ($\sigma^*$) is defined as:

$$\sigma^* = \sigma' + j\,\sigma'' \tag{6}$$

where, $\sigma'$ and $\sigma''$ are the real and imaginary parts of $\sigma^*$. The real part $\sigma'$ is derived from the measured complex impedance data:

$$\sigma' = \frac{t}{A} \times \frac{Z'}{(Z'^2 + Z''^2)} \tag{7}$$

where, $Z'$ and $Z''$ are the real and imaginary parts of the complex impedance, respectively; $A$ is area of the cross-section, and $t$ is thickness of the pellet. The frequency dependent $\sigma'$ curves, measured at several temperatures, are shown in Figures 7a–e. The conductivity values at low frequency region ($f < 10^2$ Hz) remain constant [Figures 7a–e] which are attributed to frequency-independent dc conductivity ($\sigma' \sim \sigma'_{dc}$). Whereas, at high frequency region ($f > 10^2$ Hz), the conductivity value increases with increasing frequency ($f$) suggesting frequency induced conductivity. With the increasing temperature, plateau region shifts towards a higher value of frequency for all the compounds. The contribution of dispersive component is more prominent at low temperatures. Here, we would like to point out that the effect of temperature is more predominant than that of the frequency. Each of the ac conductivity curves is fitted with the Jonscher's universal power law[50]:

$$\sigma'(T, f) = \sigma'_{dc}(T) + \xi(T)\, f^{\,n} \tag{8}$$

where, $\sigma_{dc}$ is the frequency-independent dc conductivity. The $\xi(T)$ is the pre-exponential factor, and $n$ is the frequency-exponent. Exponent $n$ ranging from 0 to 1 describes the degree of interactions between charge carriers and their surrounding environment, and $\xi$ describes the strength of polarizability of hopping ions[51]. The Jonscher's power law describes the dynamic response of charge carriers and has been used for the determination of conduction mechanism for different types of materials including ceramics, fast ionic conductors, doped ionic crystal, ion conducting glasses, disordered semiconductor and polymer[32, 34, 50, 52-54]. The value of the $\sigma'_{dc}$, obtained from the ac conductivity analysis, is found to be same as $\sigma_{dc}$ reduced from Cole-Cole plot and shown in Figure 6. The second term in Eq. (8) takes an account of the dispersion part, and provides a vital information of the conduction mechanism.

The conduction mechanism can be determined from the nature of the temperature dependent $n$ ($T$) curves. Different theoretical models for the conduction mechanism, such as, the quantum mechanical tunneling (QMT), the classical hopping over a barrier (HOB), the correlated barrier hopping (CBH), the nonoverlapping small polaron tunneling (NSPT), and the overlapping large polaron tunneling (OLPT) can be distinguished based on the nature of the temperature variation of $n$ [55-57]. In the QMT and HOB models, the value of $n$ ($T$) is independent of temperature with a constant value of ~ 0.81 and 1, respectively. For the NSPT model, the value of $n$ ($T$) increases with increasing temperature. Whereas, for the OLPT model, the value of $n$ ($T$) decreases with increasing temperature and attends a minimum value at a characteristic temperature, and then subsequently increases with increasing temperature. For the CBH model, the value of $n$ ($T$) is less than unity, and decreases with increasing temperature. According to the CBH model, the motion of charge carrier occurs through hopping between the crystal sites over the Coulomb potential barrier separating them. The height of Coulomb potential barrier is correlated with the intersite separation of charge carriers. In this model, the frequency-exponent $n$ is expressed (with first approximation) as[56]:

$$n = \frac{d(\ln\sigma')}{d\,(\ln f)} = 1 - \frac{6k_B T}{W_M} \tag{9}$$

where, $W_M$ is the effective Coulomb potential barrier. The values of the parameters $n$ and $\xi$ were determined from the fitting of the experimental ac conductivity data by Eq. (8). The temperature dependence of the parameters $n$ and $\xi$ are shown in Figures 7f and 7g, respectively. For all the compounds, the fitted values of $n$ are less than unity (1) and decrease with increasing temperature. The value of $n$ smaller than unity defines a non-zero dc conductivity. The decrease of $n$ value with increasing temperature reveals that the ionic conduction occurs through CBH. For the compound $x$ = 0, a sharp decrease in the $n$ value is observed at a characteristic temperature of $T_n$ ~ 900 K. With the increase in $x$ value, the characteristic temperature $T_n$ decreases rapidly, and attains ~ 630 K for the $x$ = 0.6 compound. Beyond $x$ = 0.6, the sharp decrease in the $n$ value occurs at a slightly higher temperature of $T_n$ ~ 720 K. The sharp decrease in the $n$ value suggests a higher ionic conduction. The decrease of $T_n$ implies a higher ionic conduction at lower temperatures for the V-substituted compounds [Figure 7f]. Furthermore, the correlated and relaxation hopping behaviour of the mobile ions can be explained by the jump relaxation model[53, 58]. In this model, the frequency-exponent $n$ can be expressed as $n = b_r/s_r$, which is the ratio of the back-hop rate ($b_r$) to the site relaxation or hole digging rate ($s_r$). The backward hop motion of a hopping ion is caused by the repulsive Coulombic interaction between the mobile ions. The site relaxation is the shift of a site potential minimum to the position of the hopping ion (after the hopping process to the site), which is caused by the rearrangements of neighboring ions[53]. With increasing the temperature, a decrease of $n$ value towards zero indicates that the site relaxation becomes faster than the backward hopping, resulting in a translation motion of the charge carriers. A decrease of $n$ (= $b_r/s_r$) value is evident with increasing the V-substitution, indicating a higher dc conduction in the forward direction. The temperature dependence of the pre-exponential factor $\xi$ reveals that there is no discernible change in $\xi$ up to 600 K. The $\xi$ value increases significantly above 600 K, reflecting a change in the strength of polarizability of hopping ions as well as hopping range in the $\beta$-phase [Figure 7g]. Furthermore, $\xi$ value increases with $x$. A higher value of $\xi$ indicates a higher carrier ionic mobility, leading to an enhanced ionic conductivity.

### 3.3.5 Electric Modulus

Electric modulus has been widely employed to study conductivity relaxation phenomenon of various compounds[34, 52]. This formalism has an advantage as it does not contain the contribution of the polarization effect from the electrode or/and mobile ion[59]. The complex form of electric modulus is represented as:

$$M^* = M' + j\,M'' = j\omega C_0\, Z^* \tag{10}$$

where, $C_0$ is the vacuum capacitance and $Z^*$ is the complex impedance. The values of real part ($M'$) and imaginary part ($M''$) of $M^*$ are derived from the measured impedance data using the following relations as:

$$M' = -Z'' \times \left(\frac{\omega A\varepsilon_0}{t}\right) \tag{11}$$

and

$$M'' = Z' \times \left(\frac{\omega A\varepsilon_0}{t}\right) \tag{12}$$

where, $\varepsilon_0$ is the free space permittivity, $\omega$ is the angular frequency, and the others symbols have their usual meaning as mentioned earlier.

The frequency dependence of the real part ($M'$) and imaginary part ($M''$) of the complex electric modulus ($M^*$) at selected temperatures are shown in Figures 8a–e and 8f–j, respectively, for all the compounds. At a given temperature, $M'$ has nearly zero value in the low frequency region and remains constant up to a characteristic frequency, and then rises suddenly and gets saturated at a higher frequency. Such frequency dependent behavior is found at all temperatures with an enhanced characteristic frequency ($f_M$) with increasing temperature.

For all the compounds, imaginary part ($M''$) of electric modulus shows a single relaxation peak at all measured temperatures. The observed peak in the $M''$ curve is broad and asymmetric in nature. Such behaviors indicate a non-Debye type relaxation process associated with ionic conduction having a distribution of relaxation time[52], as found from the impedance data. The observed peak position ($f_M$) in $M''$ curve shifts toward higher frequencies with increasing temperature revealing a decrease in relaxation time ($\tau_M$), signifying a faster movement of charge carriers. This type of behavior of $M''$ suggests the thermally activated conduction relaxation process following the hopping mechanism. The relaxation time ($\tau_M$) is determined from the experimental $M''$ data using the formula:

$$\tau_M = \frac{1}{2\pi f_M} \tag{13}$$

The temperature dependence of $\tau_M$ is shown in Figure 9d by Arrhenius plots for all the compounds. With increasing temperature ($T$), *i.e.*, decreasing $1000/T$, the value of $\tau_M$ decreases grossly. As like, the impedance data, a reversed behavior is found around the structural phase transition temperature (603–653 K for $x = 0.6$ and 603–633 K for $x = 0.8$) [Figure 9d inset].

For an ionic conductor, imaginary part ($M''$) of modulus curve can be represented by the modified Kohlrausch–Williams–Watts (KWW) function in the frequency domain. The experimentally measured $M''$ data are fitted using the modified KWW function[60] as:

$$M'' = \frac{M''_{max}}{(1-\Phi)+\left(\frac{\Phi}{1+\Phi}\right)\left[\Phi\left(\frac{f_M}{f}\right)+\left(\frac{f}{f_M}\right)^{\Phi}\right]} \tag{14}$$

where, $M''_{max}$ is the maximum value of $M''$ relaxation peak at the characteristic frequency $f_M$. The shape of each $M''$ curve in $M''$ spectra is quantified with the fitted values of $\Phi$. Temperature dependence of the $\Phi$ value is shown in Figure 9a for all the compounds. At lower temperature (~ 423 K), the values of $\Phi$ lie within the range of 0.64–0.8 for all the compounds. The deviation of $\Phi$ value from '1' reveals a non-Debye type relaxation process[34, 61]. With increasing the temperature, a sharp increase in $\Phi$ value is observed at ~ 773 K for $x = 0.4$, ~ 673 K for $x = 0.6$, ~ 623 for $x = 0.8$ and ~ 573 K for $x = 1$. Above these temperatures, the $\Phi$ attains a saturation value of ~ 0.97 close to '1', revealing the relaxation moves toward a Debye-type process. It is further evident that the effect of temperature on $\Phi$ is more dominant as the V-concentration increases. The variation of the characteristic frequency ($f_M$) with the V-concentration ($x$) is shown by the $M''/M''_{max}$ ($f$) curves [Figure 9c] at a selected temperature of 773 K in the $\beta$-phase. An increase of $f_M$ with increasing V-concentration ($x$) is evident, suggesting a decrease in $\tau_M$. Within the $\beta$-phase, a sharp decrease of the $\tau_M$ value is observed [Figures 9e–j], revealing a faster ionic conduction with increasing the V-substitution.

### 3.3.6 Dielectric Constant and Loss Factor

To estimate the contribution of local dipole-dipole relaxation, the dielectric properties have been investigated. The complex dielectric constant ($\varepsilon^*$) can be expressed as:

$$\varepsilon^* = \varepsilon' - j\,\varepsilon'' \tag{15}$$

The values of real part ($\varepsilon'$) and imaginary part ($\varepsilon''$) of $\varepsilon^*$ are derived from the measured impedance data using the following relations as:

$$\varepsilon'(\omega) = \frac{t}{\omega A\varepsilon_0} \times \frac{Z''}{(Z'^2 + Z''^2)} \tag{16}$$

and

$$\varepsilon''(\omega) = \frac{t}{\omega A\varepsilon_0} \times \frac{Z'}{(Z'^2 + Z''^2)} \tag{17}$$

The real part ($\varepsilon'$) represents the ability of the material to store electrical energy, and the imaginary part ($\varepsilon''$) represents an electrical energy loss (dissipation) in the material. The frequency responses of $\varepsilon'$ and $\varepsilon''$ are shown in Figures 10a–e and 10f–j, respectively, at selected temperatures. The $\varepsilon'$ data reveal that for a fixed temperature the value of $\varepsilon'$ decreases with increasing frequency.

The $\varepsilon''(f)$ curves decrease with increasing frequency at a fixed temperature [Figures 10f–j]. On the other hand, the value of $\varepsilon''$ increases with increasing temperature at a given frequency. The absence of any relaxation peak in the $\varepsilon''$ curves over the measured temperature and frequency ranges reveals the absence of any localized dipole relaxation in the studied compounds[62]. The complex dielectric constant can also be expressed in terms of electric modulus ($\varepsilon^* = 1/ M^*$). For a pure conduction process, a relaxation peak is typically observed in the $M''(f)$ curves, however, no peak is observed in the $\varepsilon''(f)$ curves. The appearance of relaxation peak in both $M''(f)$ and $\varepsilon''(f)$ curves corresponds to a pure localized dipole relaxation process[34, 62]. A comparison between $M''(f)$ and $\varepsilon''(f)$ curves for all the compounds ($x$ = 0–1) reveals that the relaxation peaks appear only in the $M''(f)$ curves [Figures 8f–j], not in the $\varepsilon''(f)$ curves [Figures 10f–j], confirming an absence of local dipole relaxation. Moreover, the relaxation peaks in $M''(f)$ curves and $Z''(f)$ curves appear at the same frequency values [Figure 9b], revealing that the conductivity relaxation process is associated to the oxygen ion conduction as well as long-range ionic conduction. Furthermore, the same values of frequency positions for relaxation peaks indicate the electrical homogeneity of material. At high temperatures ($\beta$-phase), value of $\varepsilon''$ increases with increasing the V-substitution, which can be related to the increase in conductivity with the substitution of V. For all temperatures, the $\varepsilon''(f)$ data are fitted by a universal power law[34, 52]:

$$\varepsilon''(T, \omega) = a(T)\,\omega^{m(T)} \tag{18}$$

where, $a(T)$ and $m(T)$ are the temperature dependent proportional constant and exponent, respectively. The value of exponent $m$ decreases with increasing temperature, and at approaches to a constant value "-1" [Figure 10p]. With increasing temperature, the decrease in $m$ value supports the thermally activated oxygen ion conduction through the CBH process, as concluded from the ac ionic conductivity study [Figure 7f].

The dielectric loss factor ($tan\ \delta$) can be determined using the following equation:

$$tan\,\delta = \frac{\varepsilon''}{\varepsilon'} = \frac{M''}{M'} = \frac{Z'}{Z''} \quad (19)$$

The frequency responses of $tan\ \delta$ are shown in Figures 10k–o, respectively, for all the compounds. Each $tan\ \delta$ curve exhibits a peak at a characteristic frequency ($f_{tan\delta}$). Overall, the $tan\ \delta$ value decreases with increasing frequency due to the frequency-activated conduction process. With increasing temperature, the value of $tan\ \delta$ increases, and $tan\ \delta$-peak shifts towards the higher frequencies. It is interesting to note that with increasing the V-substitution, $f_{tan\delta}$ shifts towards the high frequency side. The increase of $f_{tan\delta}$ with increasing the V-substitution suggests that the hopping/ jumping probability of mobile ions increases with the increasing $x$-value, *i.e.*, $V^{5+}$-substitution.

### 3.3.7 Soft-Bond Valence Sum Analysis

The oxide ion conduction pathways in $Cu_2P_{2-x}V_xO_7$ ($x$ = 0–1) have been determined by employing the soft-bond valence sum (BVS) analysis. The BVS analysis has been applied successfully to estimate the most probably oxide ion conduction pathways in oxide ion conducting materials, such as, oxide ion electrolytes[8, 63]. An empirical relationship (valence sum rule) between bond length and valence state of the ion is used to identify the possible conduction pathways[32]. The relationship is represented as:

$$V_i = \Sigma\, s_{ij} = \Sigma \exp\,[(R_0 - r_{ij})/b] \quad (20)$$

where, $V_i$ is the valence of the $i^{th}$ atom, $s_{ij}$ and $r_{ij}$ are the bond valence and bond length distance, respectively, between $i^{th}$ and $j^{th}$ atom. $R_0$ and $b$ are the empirical bond valence parameters. For the present BVS calculation, the global cutoff distance is considered to be 8 Å, so that $R_0$ and $b$ attain their constant values beyond this. Any deviation of $V_i$ from ideal valence $V_0$ indicates a dynamical disorder of O-ions. The bond valence mismatch $\Delta V = |\,V_i - V_0| = |\,V_i - 2|$ indicates the accessible volume within the structure for ionic migration. The refined crystal structural parameters [Table 1] obtained from the Rietveld analysis of the neutron diffraction patterns [Figure 1] are utilized as an input to the BVS calculation. The conduction pathways have been determined for the bond valence mismatch $\Delta V = 0.2$ valence units (v.u). The possible oxide ion migration pathways for the representative compounds with $x = 0$ (at 300 K in $\alpha$-phase and 673 K in $\beta$-phase) and $x = 1$ (at 625 K in $\beta$-phase) are shown in Figure 11. A 3D oxygen ionic conduction is evident. The major oxide ion migration pathways are along the diagonal directions of *ac* plane [Figure 11a], as well as *bc* plane [Figure 11b] and along the *b* axis [Figure 11c] of the *ab* plane. Such migration pathways are connected to form three-dimensional network within the crystal structure. The BVS analysis reveals that available volume for oxide ion conduction within the unit cell increases with increasing $x$ value. For $x = 0$, the available volume for ionic conduction at 573 and 673 K are 32.1 and 32.3 Å$^3$, respectively. Whereas, for the $x = 1$, the available volume for ionic conduction is 35.2 Å$^3$ at 625 K (estimated from the reported data in Ref. 27).

## 4. DISCUSSION

In the present study, we have demonstrated an enhancement of two orders of ionic conductivity by 50% $V^{5+}$-substitution at $P^{5+}$ site in $Cu_2P_2O_7$. Although the targeted value of conductivity (>$10^{-2}$ S•cm$^{-1}$ at ≤ 500 K) is not achieved in the present work, our results demonstrate a pathway for the enhancement of conductivity through a targeted substitution. The demonstration of the concept will be useful for the research area of SOFC. Here, we compare the ionic conductivity of the present compounds $Cu_2P_{2-x}V_xO_7$ ($x$ = 0–1) [Figure 12] with that of the other isostructural pyrophosphate compounds as well as with zirconia-based YSZ8 ($Zr_{0.92}Y_{0.16}O_{2.08}$), a commercially used electrolyte material of SOFC. The conductivity of the parent compound $Cu_2P_2O_7$ is found to be in the range of $10^{-7}$–$10^{-5}$ S cm$^{-1}$ (690–993 K) which is enhanced by the V-substitution to values in the range of $10^{-5}$–$10^{-3}$ S cm$^{-1}$ for the compound $Cu_2PVO_7$. Remarkably, the ionic conductivity values for the $x$ = 1 compound *i.e.*, $Cu_2PVO_7$ is close to that of the YSZ8 compound which is presently extensively used as an electrolyte material in commercial SOFC cells. One the other hand, when we compare the ionic conductivity value of $Cu_2P_{2-x}V_xO_7$ with that for the other reported members of the pyrophosphate family $M_2P_2O_7$, viz., $Pb_2P_2O_7$, $Zn_2P_2O_7$, $Sr_2P_2O_7$, $Ca_2P_2O_7$, $Co_2P_2O_7$, $Mn_2P_2O_7$ and even at $M$ site $Sr^{2+}$-substituted compound $SrNiP_2O_7$, it is evident that the compound with $x$ = 1 *i.e.*, $Cu_2PVO_7$ shows one of the highest conductivity. A comparable ionic conductivity value is evident only for $Co_2P_2O_7$ and $Mn_2P_2O_7$ compounds from the series. Nevertheless, our study demonstrates, for the first time, a pathway to enhance the ionic conductivity in the pyrophosphate compounds by Vanadium substitution. We anticipate that a V-ion substitution at the P site in $Co_2P_2O_7$ and $Mn_2P_2O_7$ compounds will also lead to a similar enhancement of ionic conductivity. The present crystal structural studies reveal that the substitution of larger V-ions for the P-ions increases polyhedra volume and average bond length, and subsequently an enhancement of the unit cell volume of $Cu_2P_{2-x}V_xO_7$, that increases the $(P/V)O_4$ polyhedral distortion, resulting an enhancement of ionic conductivity. Our BVS analysis also supports the enhancement of ionic conductivity with increasing the V-substitution.

Now, we shed light on additional practical implications of the studies compounds. The results of our DC transport number measurements indicate that the present $Cu_2P_{2-x}V_xO_7$ compounds have a small fraction (~ 5%) of electronic conduction in additional to the dominating ionic conductivity. Recently, Bibi *et al.*[64] have shown that such mixed ionic-electronic conductors (MIEC) can play an important role in SOFCs as electrode materials. As these materials facilitate both ionic and electronic conductions as well as simultaneous charge transfer, the overall performance of SOFC, in terms of power output, mechanical strength, and life efficiency, enhances significantly. The electronic conduction in a MIEC cathode provides the sustainable reduction of oxygen that is necessary for the efficient charge transfer for electrons as well as enabling the diffusion and transfer of oxygen ions through the electrolyte by reducing the polarization resistance. The presence of a small fraction (~ 5%) of electronic conduction in additional to the dominating ionic conductivity makes the studied compounds also suitable for new generation MIEC-cathode materials for high performance SOFC.

## 5. SUMMARY AND CONCLUSIONS

In summary, we have investigated the electrical (dc and ac conductivity, diffusivity, hopping rate, electric modulus and dielectric properties) and crystal structural properties of the polycrystalline compounds $Cu_2P_{2-x}V_xO_7$ ($x$ = 0, 0.4, 0.6, 0.8, and 1) using the impedance spectroscopy and neutron diffraction. In addition, the XPS and DC transport number measurements were carried to determine the valence states of the cations, and the relative contributions of ionic and electronic conductivities, respectively. We have successfully enhanced the ionic conductivity over two orders of magnitude with the V-substitution up to 50 atom% in $Cu_2P_{2-x}V_xO_7$. The compound with $x$ =1 shows highest conductivity among the pyrophosphates $M_2P_2O_7$ ($M$ = Zn, Pb, Sr and Ca) family. Further, the V-substitution enhances significantly the diffusivity and hopping rate of oxide ions. The observed Arrhenius behavior of the ionic conduction suggests that the conduction process is thermally activated. The BVS analysis of experimental neutron diffraction patterns highlights the 3D oxide ion conduction pathways within the crystal structure. The ac conductivity analysis reveals the ionic conduction mechanism as a correlated barrier hopping (CBH) process. In addition, electric modulus analysis indicates the faster movement of ions (long-range ionic conduction) with the V-substitution and provides an evidence for the achievement of highest conductivity. The dielectric analysis also supports the CBH process for ionic conduction and confirms an absence of any local dipole relaxations in the studied compounds. Anomalous behavior in electrical properties is found around the crystal structural phase transition temperature from $\alpha$ to $\beta$-phase for all the compounds. Although the targeted value of conductivity (>$10^{-2}$ S•cm$^{-1}$ at ≤500 K) is not achieved, the present study provides a direction to enhance the ionic conductivity by a suitable substitution of dopant without altering the charge balance at the substitution site. Our results yield a small fraction (~ 5%) of electronic conduction in additional to the dominating ionic conductivity which makes the studied compounds also suitable for new generation MIEC-cathode materials for high performance SOFC. This study also provides an in-depth understanding of the ionic conduction mechanism and associated relaxation process. Therefore, the present study paves a pathway for the enhancement of oxide ionic conductivity, which may be useful for the design and development of desired oxide-electrolyte materials for SOFCs application.

## AUTHOR INFORMATION

**Corresponding Author**

**Anup Kumar Bera** − *Solid State Physics Division, Bhabha Atomic Research Centre, Mumbai 400085, India; Homi Bhabha National Institute, Mumbai 400094, India;*
orcid.org/0000-0003-0222-0990; Email: akbera@barc.gov.in

**Authors**

**Bibhas Ghanta** − *Solid State Physics Division, Bhabha Atomic Research Centre, Mumbai 400085, India; Homi Bhabha National Institute, Mumbai 400094, India;*
orcid.org/0009-0002-6782-4488;

**Kuldeep Singh Chikara** − *Solid State Physics Division, Bhabha Atomic Research Centre, Mumbai 400085, India; Homi Bhabha National Institute, Mumbai 400094, India;*
orcid.org/0009-0002-0748-1516;

**Uttam Kumar Goutam** - *Technical Physics Division, Bhabha Atomic Research Centre, Mumbai 400085, India;*
orcid.org/0000-0002-8789-3547;

**Seikh Mohammad Yusuf** − *Solid State Physics Division, Bhabha Atomic Research Centre, Mumbai 400085, India; Homi Bhabha National Institute, Mumbai 400094, India;*
orcid.org/0000-0003-1898-4610;

**Funding Source**

S. M. Y. acknowledges the financial assistance from ANRF, Department of Science and Technology, Government of India under the J. C. Bose fellowship program (Ref. No. JCB/2023/000014).

**Notes**

The authors declare no competing financial interest.

# FIGURES

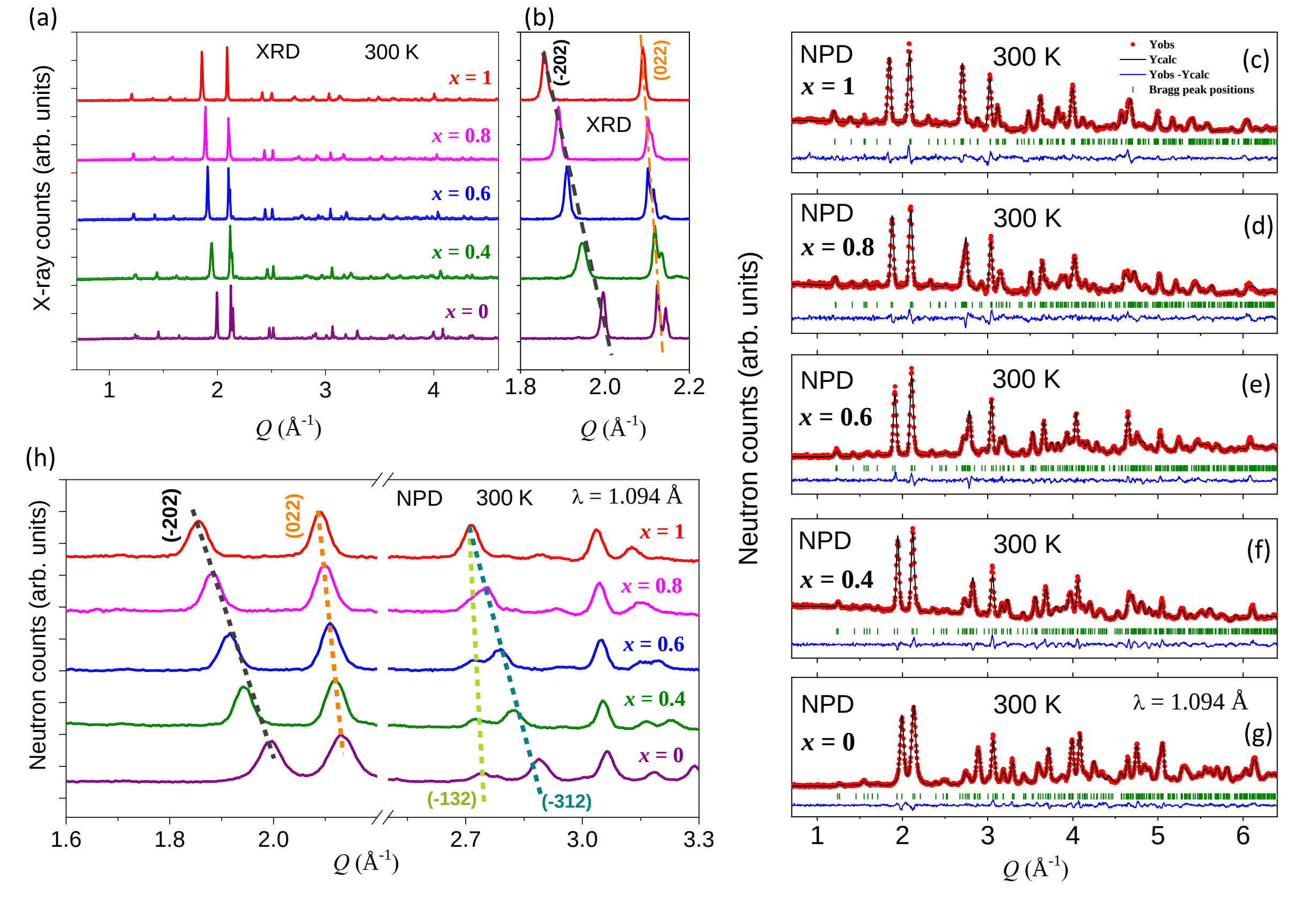

**Figure 1.** (a) As measured x-ray diffraction (XRD) patterns of $Cu_2P_{2-x}V_xO_7$ ($x$ = 0, 0.4, 0.6, 0.8 and 1) at 300 K. (b) Enlarged view of XRD patterns over a selected momentum transfer range of 1.8–2.2 Å$^{-1}$ to illustrate the Bragg peaks (-202) and (022). The dotted lines are guide to the eye. (c–g) Rietveld analyzed neutron powder diffraction patterns for all the compounds at 300 K. The observed and calculated diffraction patterns are shown by the filled circles (red) and solid lines (black), respectively. The difference between observed and calculated patterns is shown by the thin line (blue) at the bottom of each panel. The vertical bars are the allowed Bragg peak positions. (h) The selected region of neutron diffraction patterns over a momentum transfer range of 1.6–3.3 Å$^{-1}$ highlights the Bragg peaks (-202), (022), (-132) and (-312). The dotted lines are guide to the eye.

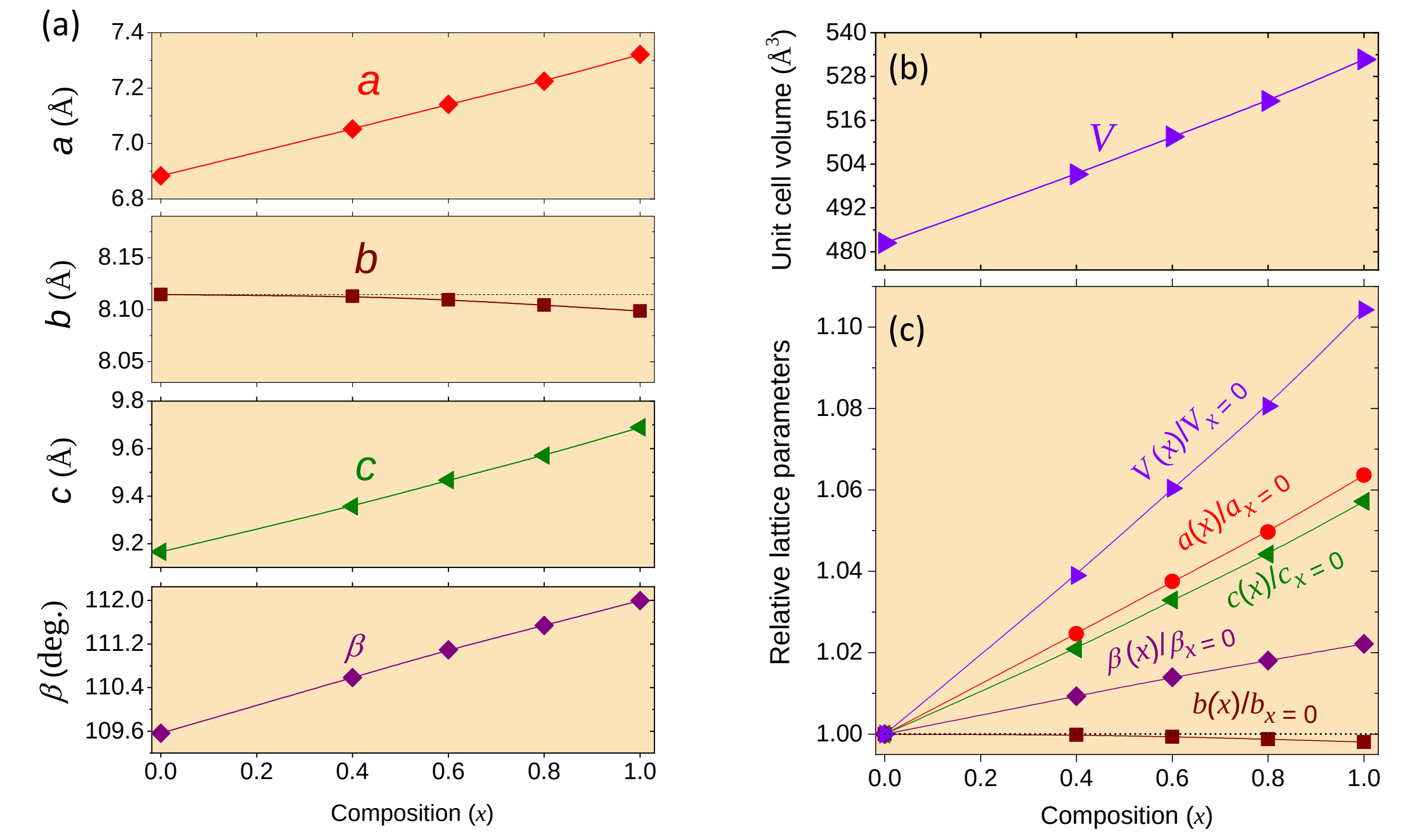

**Figure 2.** The Vanadium concentration ($x$) dependent variation of (a) lattice constants ($a$, $b$ and $c$) and monoclinic angle $\beta$, and (b) unit cell volume ($V$) at 300 K for $Cu_2P_{2-x}V_xO_7$ ($x$ = 0, 0.4, 0.6, 0.8 and 1). (c) The relative changes of $a$, $c$, $b$, $\beta$ and $V$ with respect to $x$ = 0 compound as a function of Vanadium concentration ($x$). The dotted lines are guide to the eye, highlighting the slightly reduction of $b$ with increasing Vanadium concentration ($x$).

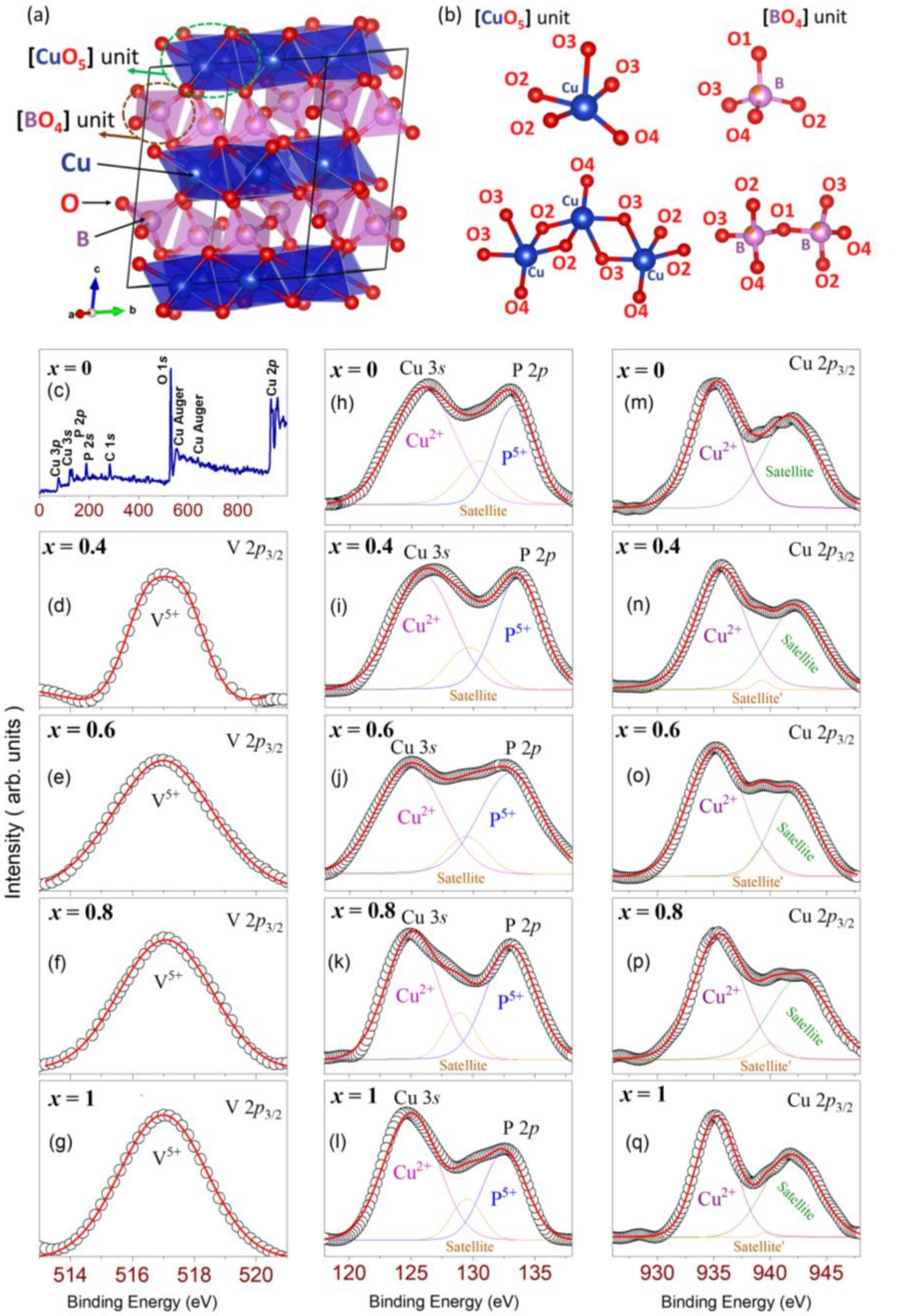

**Figure 3.** (a) A schematic crystal structure (space group: $C2/c$) of $Cu_2P_{2-x}V_xO_7$ ($x$ = 0, 0.4, 0.6, 0.8 and 1). The $CuO_5$ and tetrahedral $BO_4$ units ($B$ = P/V) are shown by blue and pink colours, respectively, and highlighted by dashes lines. The thin black lines show the dimension of the unit cell. The atomic fractions of P and V at the $B$ site are shown by pink and orange parts of the spheres, respectively. (b) The different Cu−O and $B$−O bonds in $CuO_5$ and $BO_4$ units, respectively. In *ab* plane, $CuO_5$ units make a zigzag chain by sharing their O2-O2 and O3-O3 edges of $CuO_5$ unit. Two $BO_4$ units are connected by sharing O1 atom at the vertex of $BO_4$ unit. (c) XPS survey spectrum for the compound $x$ = 0. (d−g) XPS core level spectra of V $2p_{3/2}$ for the compounds $x$ = 0.4, 0.6, 0.8 and 1. (h−l) XPS core level spectra of Cu 3$s$ and P 2$p$, and (m−q) XPS core level spectra of Cu $2p_{3/2}$ for the compounds $x$ = 0, 0.4, 0.6, 0.8 and 1.

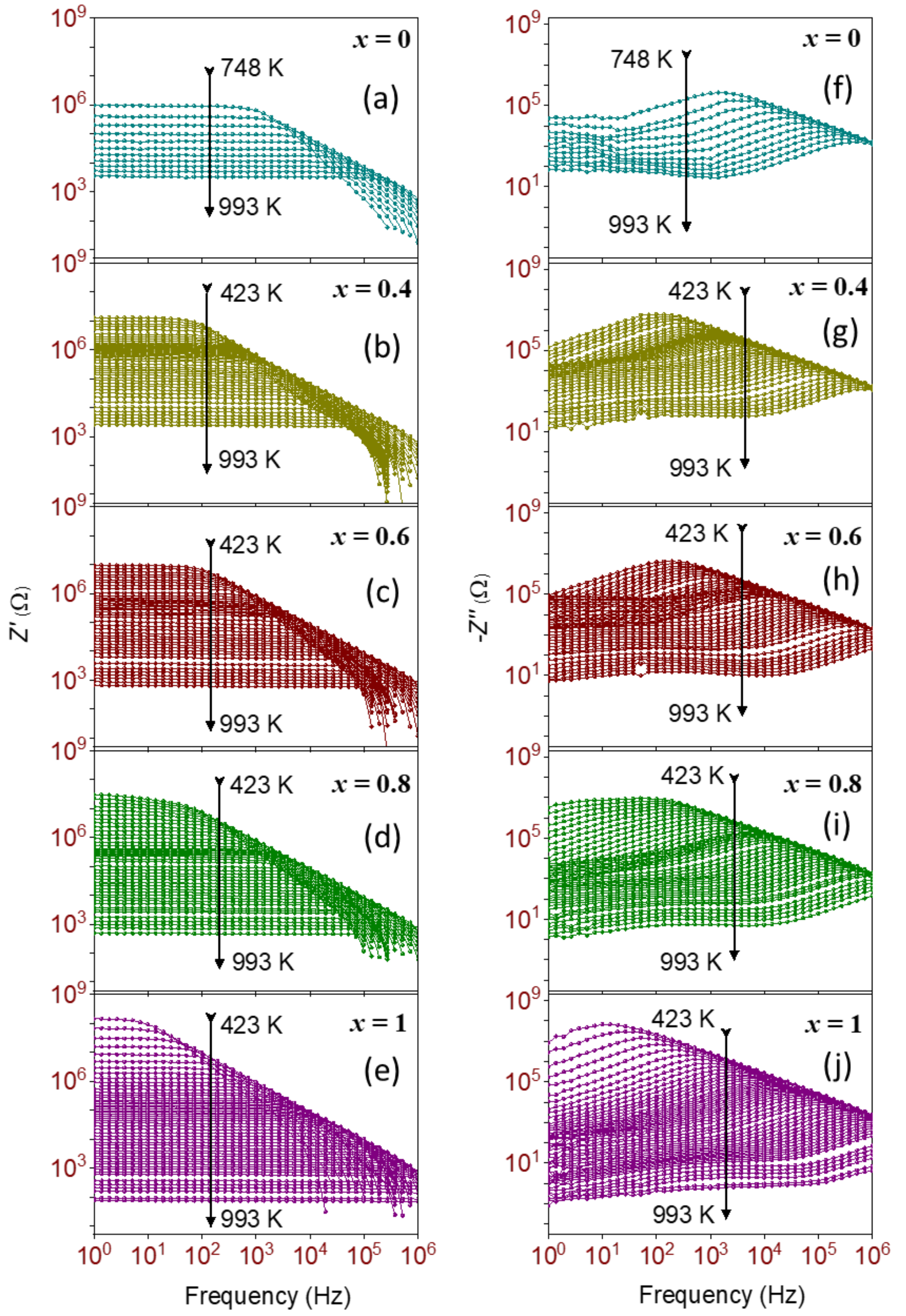

**Figure 4.** Frequency dependent (a–e) real part ($Z'$) and (f–j) imaginary part ($Z''$) of complex impedance ($Z^*$) of $Cu_2P_{2-x}V_xO_7$ ($x$ = 0, 0.4, 0.6, 0.8 and 1) at selected temperatures.

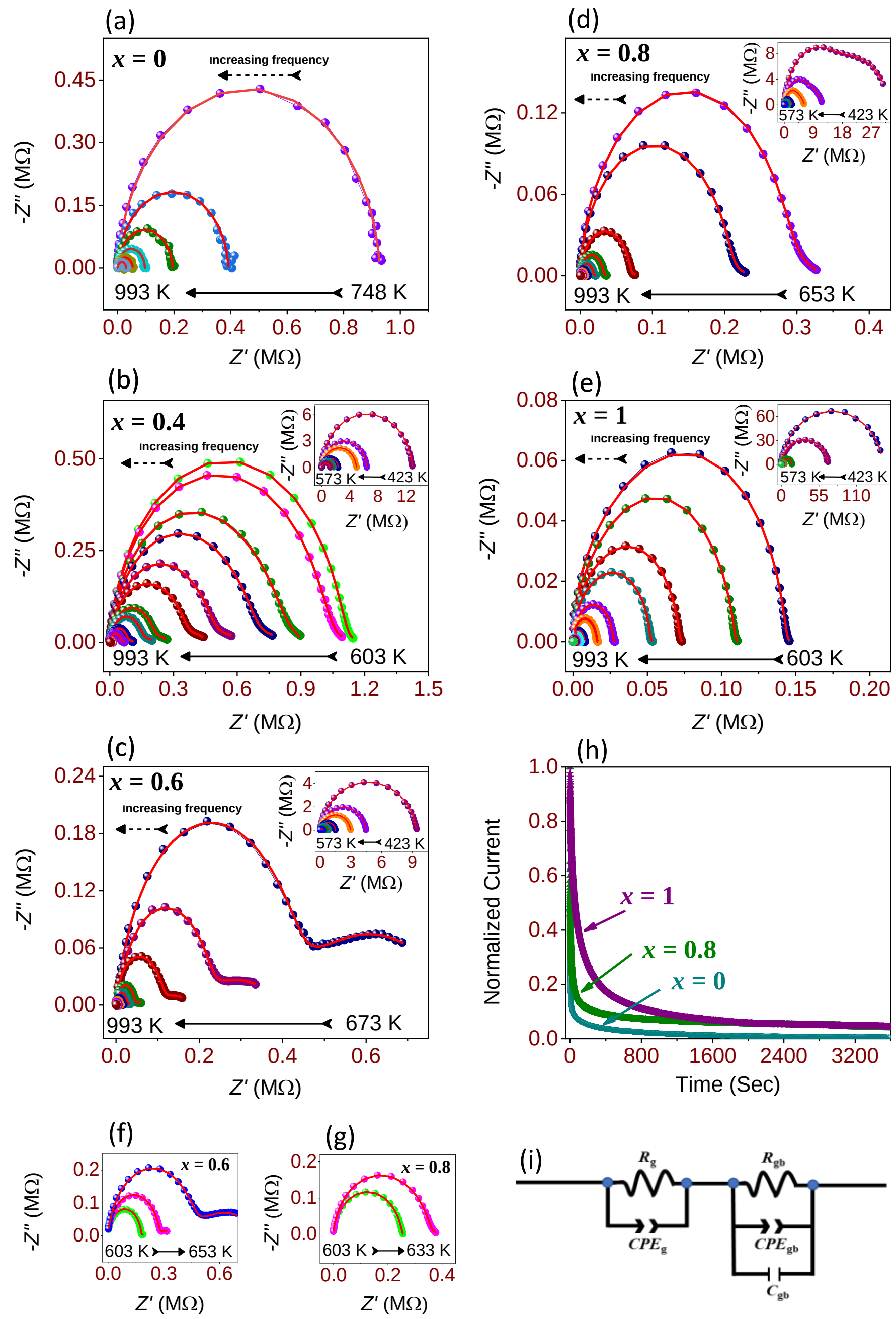

**Figure 5.** (a−g) The Cole-Cole plots of complex impedance for $Cu_2P_{2-x}V_xO_7$ ($x$ = 0, 0.4, 0.6, 0.8 and 1) at selected temperatures. The lines passing through the experimental data points are fitted curves using the equivalent circuit model as shown in (i). The insets show the Cole−Cole plots at low-temperatures. (f and g) The Cole−Cole plots around the crystal structural phase transition temperature regions. (i) The schematic diagram of equivalent circuit model comprised of grain and grain boundary contributions. (h) The normalized current vs time curve for the compounds $x$ = 0, 0.8 and 1.

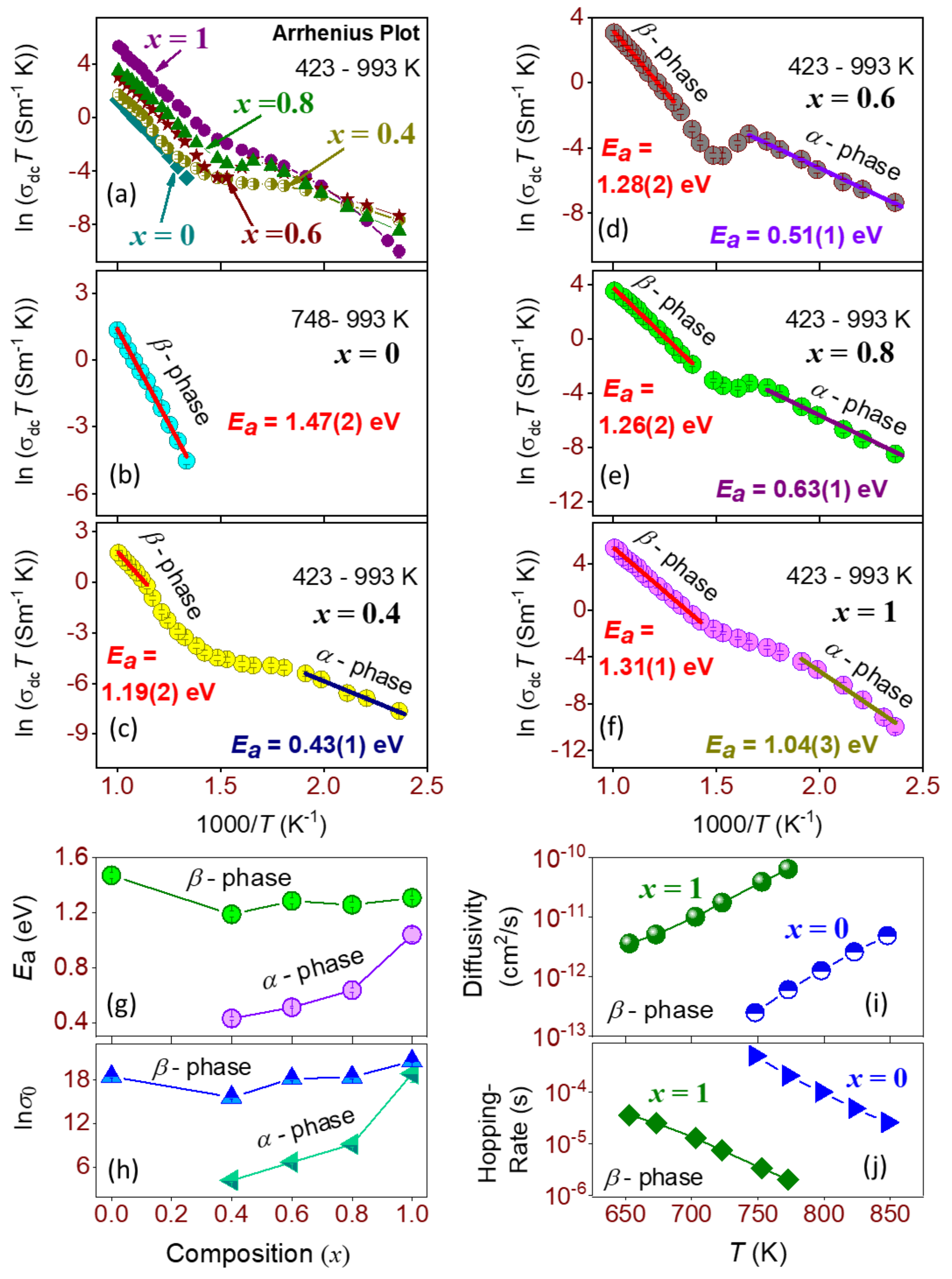

**Figure 6.** (a) Temperature dependent of total dc conductivity ($\sigma_{dc}$) for $Cu_2P_{2-x}V_xO_7$ ($x$ = 0, 0.4, 0.6, 0.8 and 1). (b–f) Temperature dependent $\sigma_{dc}$ (Arrhenius plots) for individual compounds. The solid straight lines through the data points are the linear fits to the $\sigma_{dc}$ data. (g) and (h) The variation of activation energy ($E_a$) and pre-exponential factor ($\sigma_0$), respectively, with Vanadium-concentration ($x$). (i) and (j) The temperature dependent variation of diffusion constant and hopping rates of oxide ions, respectively, in the $\beta$-phase for the two end compounds with $x$ = 0 and 1.

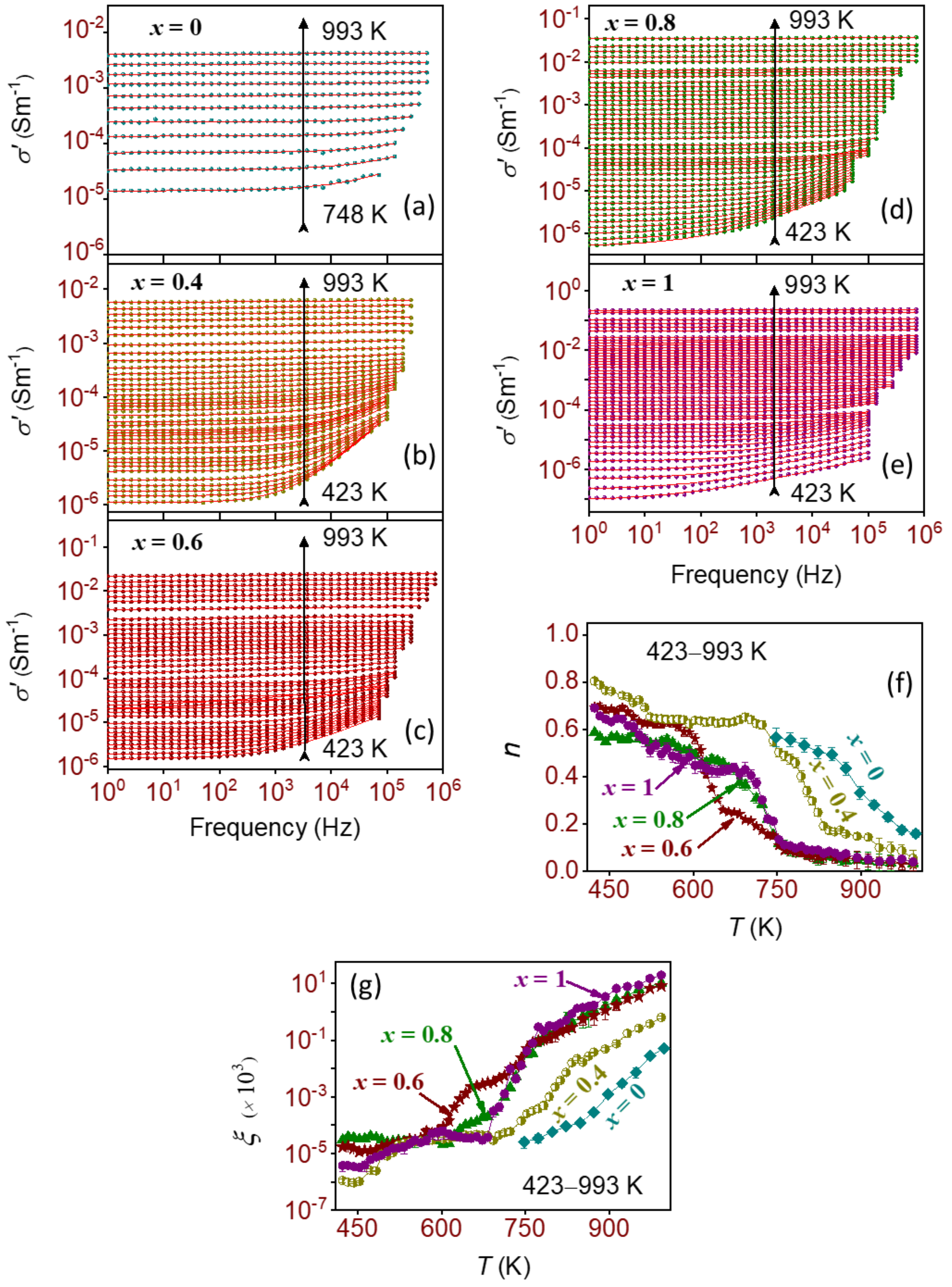

**Figure 7.** (a–e) The ac conductivity spectra as a function of frequency for $Cu_2P_{2-x}V_xO_7$ ($x$ = 0, 0.4, 0.6, 0.8 and 1) at selected temperatures, respectively. The solid lines (red) through the data points are fitted curves using the Jonscher's power law [eq 8]. (f) and (g) The variation of frequency-exponent ($n$) and pre-exponential factor ($\xi$) as a function of temperature, respectively.

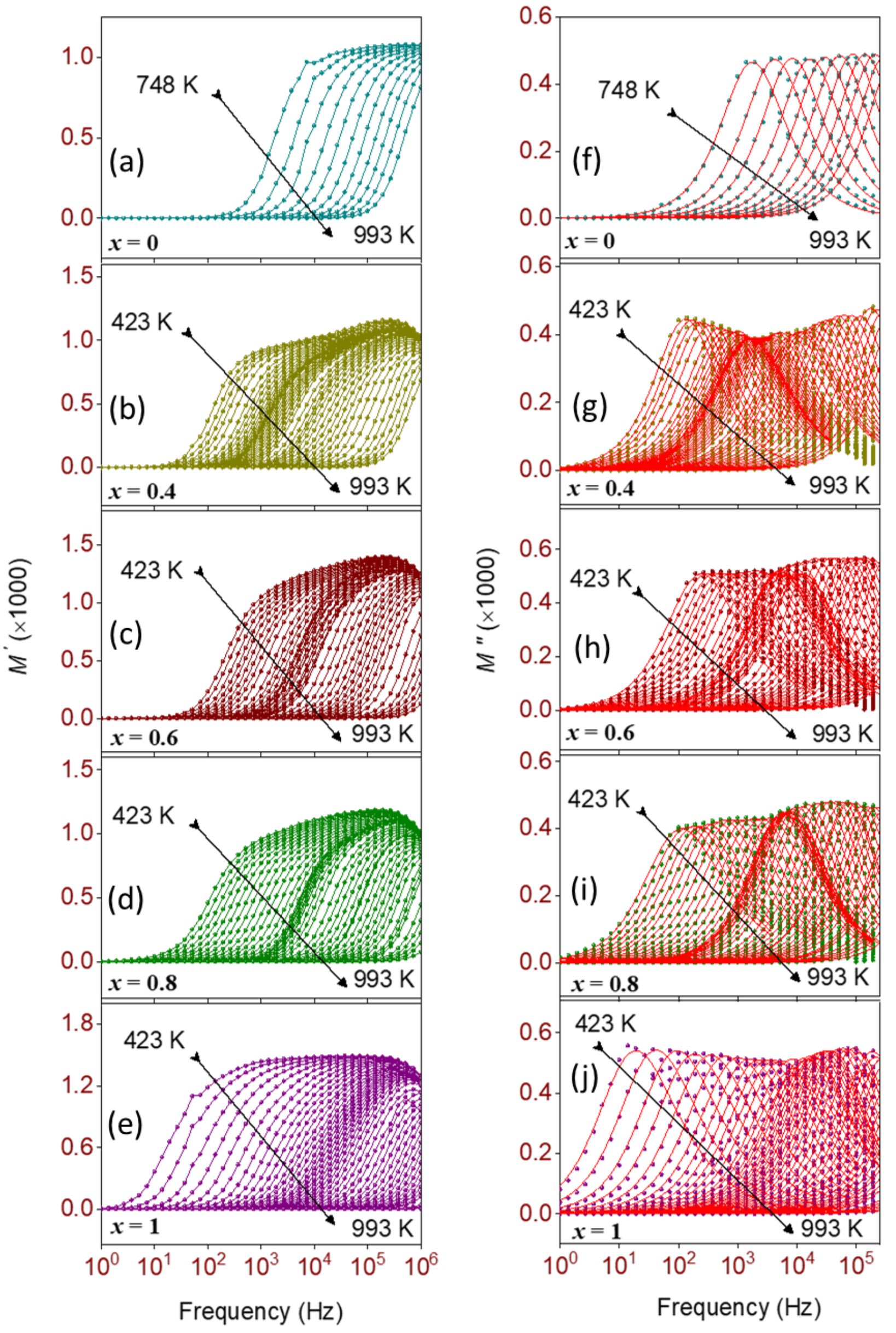

**Figure 8.** Frequency dependent (a–e) real part ($M'$) and (f–j) imaginary part ($M''$) of complex modulus ($M^*$) for $Cu_2P_{2-x}V_xO_7$ ($x$ = 0, 0.4, 0.6, 0.8 and 1) at selected temperatures. The solid lines (red) through the data points in (f–j) are the fitted curves using the modified KWW function [eq 14].

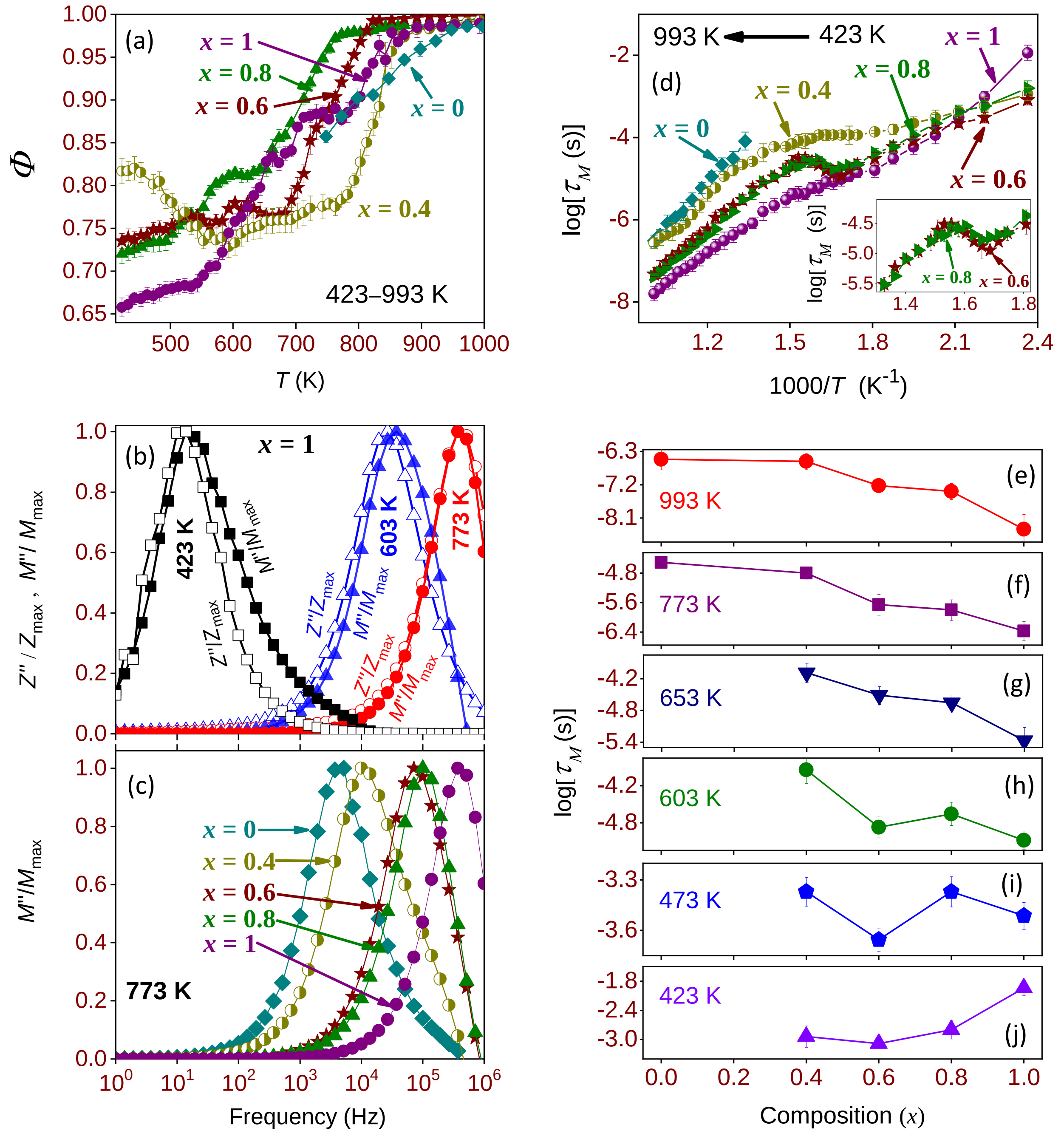

**Figure 9.** (a) The temperature variation of KWW stretching parameter ($\Phi$) for $Cu_2P_{2-x}V_xO_7$ ($x$ = 0, 0.4, 0.6, 0.8 and 1). (b) Normalized imaginary parts of impedance ($Z''/Z''_{max}$) and electric modulus ($M''/M''_{max}$) as a function of frequency for $x$ = 1 shown for three re-presentive temperatures 423, 603, and 773 K. (c) Normalized imaginary parts of electric modulus ($M''/M''_{max}$) as a function of frequency at 773 K for all the compounds ($x$ = 0, 0.4, 0.6, 0.8 and 1). (d) The temperature dependent relaxation time ($\tau_M$) for all the compounds ($x$ = 0, 0.4, 0.6, 0.8 and 1). The inset shows a zoomed view around the crystal structural phase transition temperature from $\alpha$ to $\beta$-phase. (e–j) The variation of relaxation time with composition at several selected temperatures.

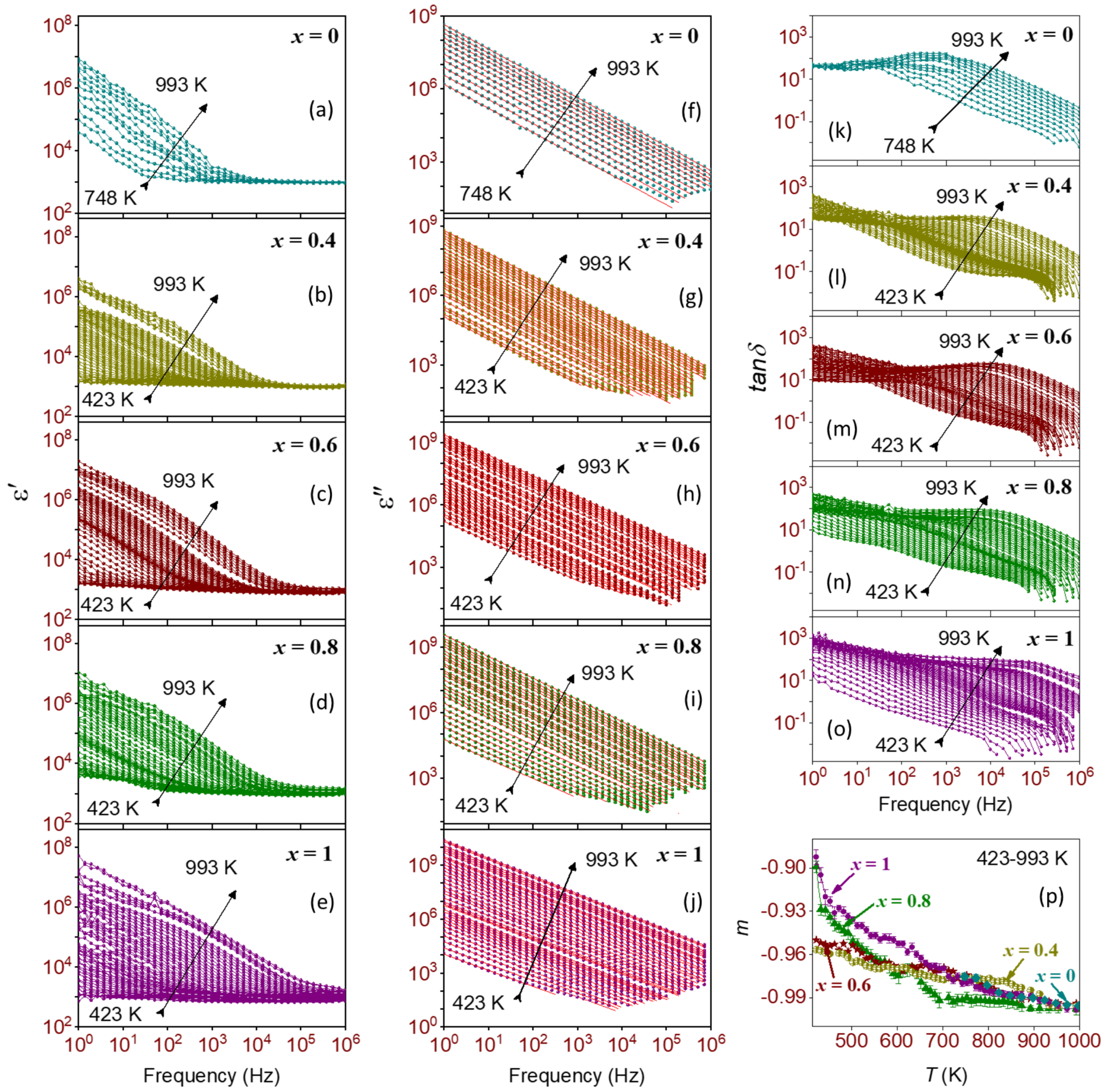

**Figure 10.** Frequency dependent (a–e) real part ($\varepsilon'$) and (f–j) imaginary part ($\varepsilon''$) of complex dielectric constant ($\varepsilon^*$) for $Cu_2P_{2-x}V_xO_7$ ($x$ = 0, 0.4, 0.6, 0.8 and 1) at selected temperatures. The solid lines (red) through the data points in (f–j) are the fitted curves using the universal power law [eq 18]. (k–o) Frequency dependent dielectric loss factor at selected temperatures for $Cu_2P_{2-x}V_xO_7$ ($x$ = 0, 0.4, 0.6, 0.8 and 1). (p) The temperature dependence of exponent $m$ for all the compounds $x$ = 0−1.

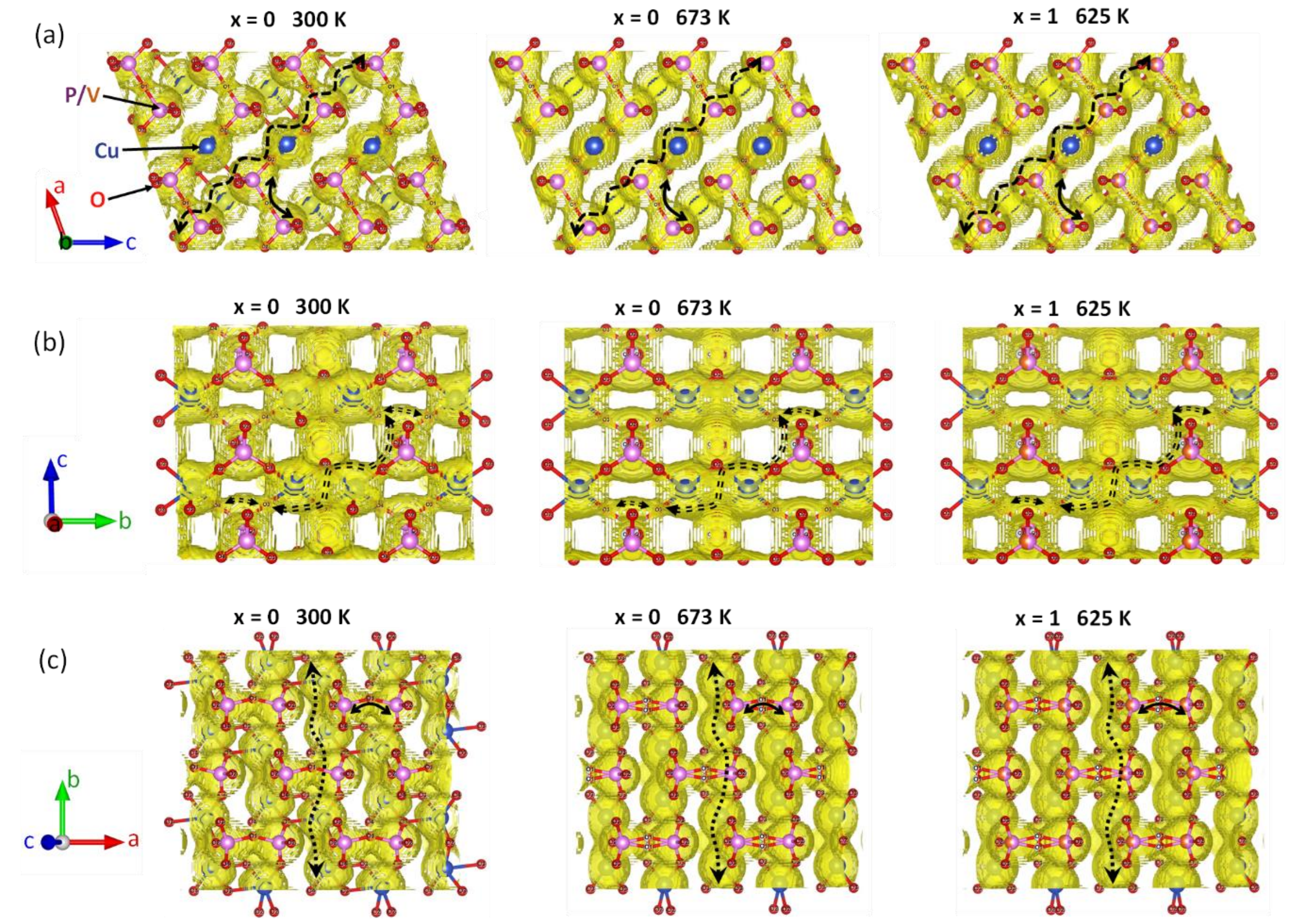

**Figure 11.** The projections of oxygen ions migration pathways through the polyhedral network of the compounds $Cu_2P_{2-x}V_xO_7$ with $x$ = 0 [at 300 K ($\alpha$-phase) and 673 K ($\beta$-phase)], and $x$ = 1 [at 625 K ($\beta$-phase)] in the (a) *ac*, (b) *bc* and (c) *ab* plane, respectively. The dotted and solid arrows indicate the most probable oxide ion migration directions.

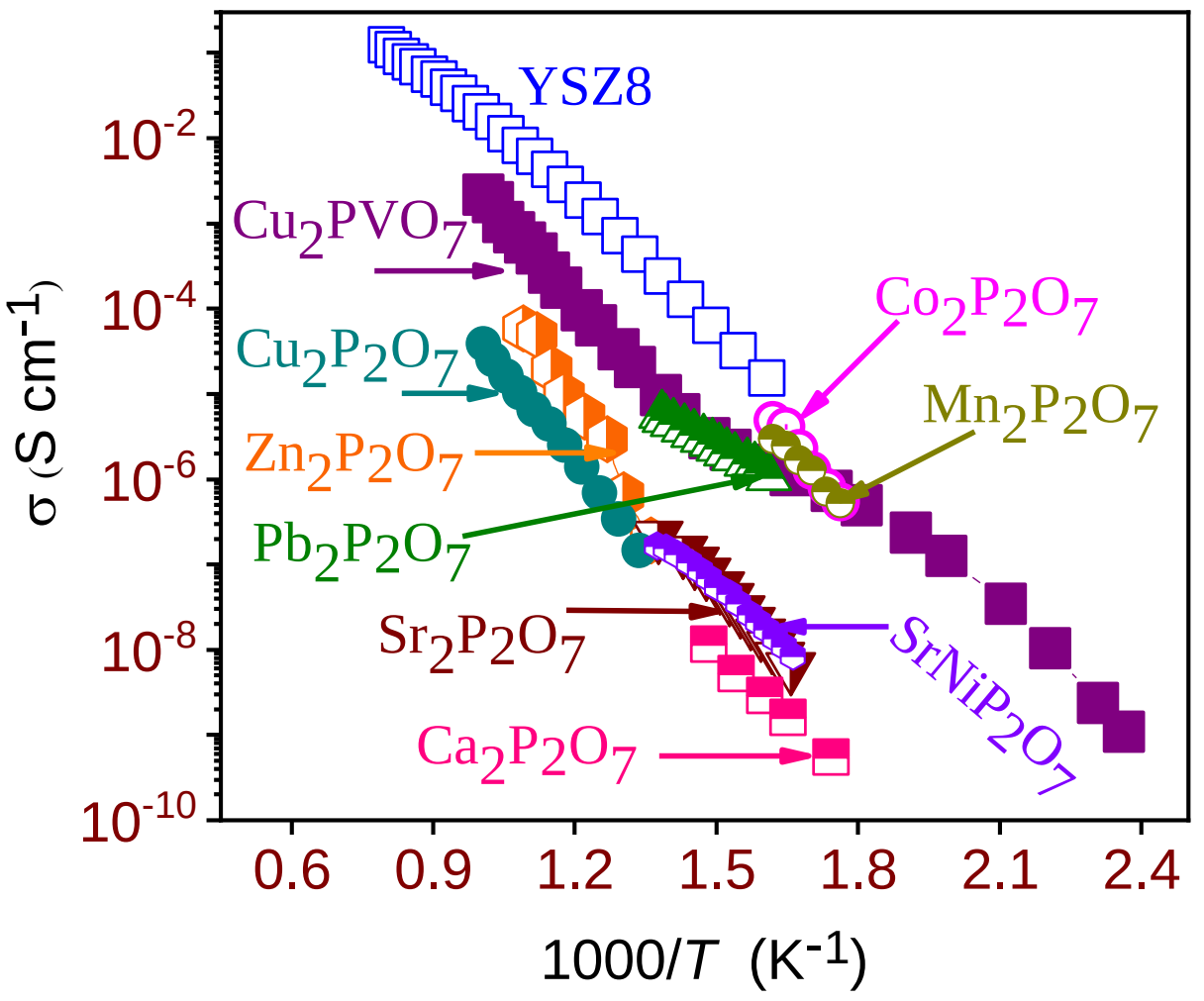

**Figure 12.** Comparison of bulk ionic conductivity of $Cu_2PVO_7$ with some of the reported ionic conductors, such as, YSZ8[65] ($Zr_{0.92}Y_{0.16}O_{2.08}$), and pyrophosphate compounds viz., $Cu_2P_2O_7$ (the present work), $Pb_2P_2O_7$[9], $Zn_2P_2O_7$[10], $Sr_2P_2O_7$[12], $Ca_2P_2O_7$[66], $Co_2P_2O_7$[67], $Mn_2P_2O_7$[67] and $SrNiP_2O_7$[68].

**Table 1.** The Rietveld-Refined Lattice Parameters, Unit Cell Volumes and Fractional Atomic Coordinates for the Compounds $Cu_2P_{2-x}V_xO_7$ ($x$ = 0, 0.4, 0.6, 0.8 and 1) at 300 K.

| | Site | $x$ = 0 | $x$ = 0.4 | $x$ = 0.6 | $x$ = 0.8 | $x$ = 1 |
|---|---|---|---|---|---|---|
| $a$ (Å) | | 6.883(1) | 7.053(4) | 7.141(2) | 7.225(3) | 7.321(2) |
| $b$ (Å) | | 8.115(2) | 8.113(2) | 8.109(3) | 8.104(3) | 8.099(2) |
| $c$ (Å) | | 9.166(3) | 9.357(6) | 9.467(4) | 9.572(4) | 9.689(3) |
| $\beta$ (°) | | 109.56(2) | 110.59(3) | 111.09(2) | 111.54(1) | 111.99(2) |
| $V$ (Å$^3$) | | 482.39(5) | 501.21(6) | 511.55(3) | 521.28(5) | 532.70(3) |
| | | | | | | |
| Cu | 8*f* | | | | | |
| *x/a* | | 0.9851(8) | 0.9777(4) | 0.9779(6) | 0.9706(4) | 0.9716(5) |
| *y/b* | | 0.3152(6) | 0.3150(6) | 0.3142(6) | 0.3127(8) | 0.3141(3) |
| *z/c* | | 0.5085(8) | 0.5083(8) | 0.5091(4) | 0.5104(5) | 0.5130(4) |
| Occ. | | 1 | 1 | 1 | 1 | 1 |
| | | | | | | |
| P/V | 8*f* | | | | | |
| *x/a* | | 0.2022(8) | 0.1970(6) | 0.1998(5) | 0.2037(5) | 0.2003(3) |
| *y/b* | | 0.0045(8) | 0.0133(6) | 0.0094(5) | 0.0111(4) | 0.0045(3) |
| *z/c* | | 0.2074(7) | 0.2067(7) | 0.2108(4) | 0.2135(4) | 0.2113(5) |
| Occ. | | 1/0.0 | 0.8/0.2 | 0.7/0.3 | 0.6/0.4 | 0.5/0.5 |
| | | | | | | |
| O1 | 4*e* | | | | | |
| *x/a* | | 0 | 0 | 0 | 0 | 0 |
| *y/b* | | 0.0500(1) | 0.0499(3) | 0.0539(2) | 0.0602(7) | 0.0734(2) |
| *z/c* | | 0.25 | 0.25 | 0.25 | 0.25 | 0.25 |
| Occ. | | 1 | 1 | 1 | 1 | 1 |
| | | | | | | |
| O2 | 8*f* | | | | | |
| *x/a* | | 0.3734(9) | 0.3803(3) | 0.3828(4) | 0.3824(5) | 0.3893(7) |
| *y/b* | | 0.9988(6) | 0.9985(4) | 0.9969(6) | 0.9992(6) | 0.9995(4) |
| *z/c* | | 0.3629(6) | 0.3613(6) | 0.3652(6) | 0.3652(8) | 0.3650(2) |
| Occ. | | 1 | 1 | 1 | 1 | 1 |
| | | | | | | |
| O3 | 8*f* | | | | | |
| *x/a* | | 0.2196(9) | 0.2206(2) | 0.2214(4) | 0.2217(4) | 0.2193(6) |
| *y/b* | | 0.1524(8) | 0.1560(5) | 0.1564(8) | 0.1576(4) | 0.1602(5) |
| *z/c* | | 0.1117(8) | 0.1076(9) | 0.1083(8) | 0.1074(7) | 0.1044(4) |
| Occ. | | 1 | 1 | 1 | 1 | 1 |
| O4 | 8*f* | | | | | |
| *x/a* | | 0.1734(6) | 0.1670(2) | 0.1656(2) | 0.1671(6) | 0.1673(3) |
| *y/b* | | 0.8475(9) | 0.8498(6) | 0.8491(9) | 0.8450(4) | 0.8496(5) |
| *z/c* | | 0.1160(8) | 0.1171(8) | 0.1189(5) | 0.1193(4) | 0.1206(5) |
| Occ. | | 1 | 1 | 1 | 1 | 1 |

**Table 2.** The Values of Activation Energy ($E_a$) and Pre-exponential Factor ($\sigma_0$) for the Compounds $Cu_2P_{2-x}V_x O_7$ ($x$ = 0, 0.4, 0.6, 0.8 and 1) which are Determined from the Arrhenius Relation for the $\alpha$ and $\beta$ phases.

| | $\beta$-phase | | $\alpha$-phase | |
|---|---|---|---|---|
| Compound | $E_a$ (eV) | ln$\sigma_0$ | $E_a$ (eV) | ln$\sigma_0$ |
| $x$ = 0 | 1.47(2) | 18.48(2) | | |
| $x$ = 0.4 | 1.19(2) | 15.63(2) | 0.43(1) | 4.12(1) |
| $x$ = 0.6 | 1.28(2) | 18.15(2) | 0.51(1) | 6.64(2) |
| $x$ = 0.8 | 1.26(2) | 18.39(2) | 0.63(1) | 9.12(1) |
| $x$ = 1 | 1.31(1) | 20.60(1) | 1.04(3) | 18.87(4) |